\documentclass[11pt]{article}
\pdfoutput=1 

\usepackage{jheppub} 

\usepackage[T1]{fontenc} 
\graphicspath{{./figures/}}
\usepackage{bm}
\usepackage{dcolumn}
\usepackage{color}
\usepackage{tabularx}
\usepackage{subfig}
\usepackage{mathtools}
\usepackage{braket}
\usepackage{multirow}
\usepackage{amsmath, amssymb, amsfonts}
\newcommand{\dnu}{\nu}

\title{\boldmath Color superconductivity on the lattice
  --- analytic predictions from QCD in a small box}

\author[a,b]{Takeru Yokota,}
\author[c]{Yuta Ito,}
\author[d,e]{Hideo Matsufuru,}
\author[f]{Yusuke Namekawa,}
\author[d,g]{Jun Nishimura,}
\author[h]{Asato Tsuchiya,}
\author[i]{and Shoichiro Tsutsui}

\affiliation[a]{Interdisciplinary Theoretical and Mathematical Sciences Program (iTHEMS), RIKEN, Wako, Saitama 351-0198, Japan}
\affiliation[b]{Institute for Solid State Physics, The University of Tokyo, Kashiwa, Chiba 277-8581, Japan}
\affiliation[c]{National Institute of Technology, Tokuyama College, Gakuendai, Shunan, Yamaguchi 745-8585, Japan}
\affiliation[d]{High Energy Accelerator Research Organization (KEK), 1-1 Oho, Tsukuba, Ibaraki 305-0801, Japan}
\affiliation[e]{Department of Accelerator Science, School of High Energy Accelerator Science, Graduate University for Advanced Studies (SOKENDAI), 1-1 Oho, Tsukuba, Ibaraki 305-0801, Japan}
\affiliation[f]{Education and Research Center for Artificial Intelligence and Data Innovation, Hiroshima University, 1-1-89 Higashisendamachi, Naka Ward, Hiroshima, 730-0053, Japan}
\affiliation[g]{Department of Particle and Nuclear Physics, School of High Energy Accelerator Science, Graduate University for Advanced Studies (SOKENDAI), 1-1 Oho, Tsukuba, Ibaraki 305-0801, Japan}
\affiliation[h]{Department of Physics, Shizuoka University, 836 Ohya, Suruga-ku, Shizuoka 422-8529, Japan}
\affiliation[i]{Theoretical Research Division, Nishina Center, RIKEN, Wako, Saitama 351-0198, Japan}

\emailAdd{takeru.yokota@riken.jp}
\emailAdd{y-itou@tokuyama.ac.jp}
\emailAdd{hideo.matsufuru@kek.jp}
\emailAdd{namekawa@hiroshima-u.ac.jp}
\emailAdd{jnishi@post.kek.jp}
\emailAdd{tsuchiya.asato@shizuoka.ac.jp}
\emailAdd{shoichiro.tsutsui@riken.jp}

\preprint{RIKEN-iTHEMS-Report-23, KEK-TH-2496}

\abstract{We investigate color superconductivity on the lattice
  using the gap equation for the Cooper pair condensate.
  The weak coupling analysis is justified by choosing the physical size of the lattice to be smaller than the QCD scale, while keeping the aspect ratio of the lattice small enough to suppress thermal excitations.
  In the vicinity of the critical coupling constant
  that separates the superconducting phase and the normal phase,
  the gap equation can be linearized, and by solving the corresponding
  eigenvalue problem,
  we obtain the critical point and the Cooper pair condensate without assuming its
  explicit form.
  The momentum components of the condensate suggest
  spatially isotropic s-wave superconductivity
  with Cooper pairs formed by quarks near the Fermi surface.
  The chiral symmetry in the massless limit is spontaneously broken
  by the Cooper pair condensate, which turns out to be
  dominated by the scalar and the pseudo-scalar components.
  Our results provide useful predictions, in particular, for
  future lattice simulations
  based on methods to overcome the sign problem
  such as the complex Langevin method.
}

\begin{document} 
\maketitle
\flushbottom

\section{Introduction}
Elucidating the nature of quark matter is one of the long-standing issues in modern physics.
In fact, the phase structure of QCD at finite quark density
is expected to be extremely rich
and its exploration has been challenged on both experimental and theoretical sides.
The experimental attempts include the heavy-ion collision \cite{blaschke_probing_2022}
and the observation of neutron stars by X-ray and gravitational waves \cite{orsaria_phase_2019}.
On the theoretical side, the access to the finite density region based on the underlying theory, QCD,
has been restricted due to the notorious sign problem,
which represents the breakdown of importance sampling in lattice Monte Carlo simulations; see, e.g., Ref.~\cite{Nagata:2021ugx}.
However, the situation is changing
drastically
thanks to the recent development of various approaches
such as the complex Langevin
method \cite{par83,kla84,Aarts:2009uq,Aarts:2011ax,Nagata:2015uga,Nagata:2016vkn},
the Lefschetz thimble
method \cite{witten_analytic_2010,Cristoforetti:2012su,Cristoforetti:2013wha,Fujii:2013sra,Alexandru:2015sua,Fukuma:2017fjq,Fukuma:2020fez,Fukuma:2021aoo,Fujisawa:2021hxh},
the path optimization
method \cite{mori_toward_2017,mori_application_2018,alexandru_finite-density_2018} and the tensor
renormalization group method \cite{levin_tensor_2007}. In particular, the complex Langevin method
has been successfully applied to QCD at finite
density \cite{sexty_simulating_2014,Aarts:2014bwa,Fodor:2015doa,nagata_complex_2018,kogut_applying_2019,sexty_calculating_2019,scherzer_deconfinement_2020,ito_complex_2020,Attanasio:2022mjd,tsutsui_color_2022,namekawa_flavor_2022}.

One of the most intriguing phenomena in QCD at finite density
is the color superconductivity (CSC)
due to the formation of Cooper
pairs by quarks \cite{bar77,frautschi_asymptotic_1980, bailin_superfluidity_1981, alford_qcd_1998}.
This phenomenon is predicted from the calculation of a one-gluon exchange diagram,
which gives rise to an attractive force in the color anti-triplet channel of quark pairs causing
the Cooper instability at low temperatures.
Such weak coupling calculations are justified, for instance,
when the chemical potential is sufficiently larger than
the QCD scale $\Lambda_{\rm QCD}\sim 200~\mathrm{MeV}$
due to the asymptotic freedom.
However, this setup is
not easy to realize in lattice simulations
since it requires the lattice spacing to be sufficiently smaller than the inverse of the chemical potential and one typically has to use a huge lattice
to suppress finite size effects.

Another possibility for validating the weak coupling calculations
is to consider QCD in a box
which is sufficiently smaller than $\Lambda_{\rm QCD}^{-1} \sim 1~\mathrm{fm}$.
This setup is particularly useful
in testing the aforementioned methods
for finite density QCD
since the required lattice size is quite modest.
For instance,
two-color QCD\footnote{This is an SU(2) gauge theory with fermions in the
  fundamental representation, which does not suffer from the sign problem
even at finite density.}
in a small box was investigated by lattice simulations
at finite density
\cite{hands_numerical_2010}.
Related perturbative calculations have been done
in QCD on $S^3 \times S^1$ at one loop \cite{hands_qcd_2010},
where 
the quark number susceptibility, 
the Polyakov line and the chiral condensate
are found to have intriguing dependence on the chemical potential.
There are also lattice simulations of the Nambu--Jona-Lasinio (NJL) model at finite density, which exhibit some evidence for BCS diquark condensation \cite{Hands:2002mr}.

Although superconductivity is basically a weak coupling phenomenon,
the analysis of the Cooper pair condensate requires some methods to sum up
infinitely many loop diagrams as in the analysis of
the chiral condensate in the NJL model.
An established tool for that is the gap equation, which is a
self-consistent equation for the
Cooper pair condensate that can be derived from the Dyson equation.
It is usually formulated in the continuum to obtain useful information, for instance,
on how the energy gap scales with the coupling constant; see Ref.~\cite{alf08} and references therein.
In order to make quantitative predictions, however,
further simplification using an assumption on the form of the
Cooper pair condensate
is needed since the gap equation is too complicated to be solved in full generality.

In this paper we study the CSC
on a finite lattice
so that the gap equation becomes a finite number of coupled equations.
Furthermore we focus on the vicinity of the critical point, where the energy gap is small assuming a continuous phase transition, so that the gap equation reduces to a linear equation.
By simply solving an eigenvalue problem associated with this linear equation,
we can investigate the existence of a non-trivial solution
and make quantitative predictions on the CSC.
In particular,
no assumption on the form of the Cooper pair condensate
is required unlike
similar calculations in the continuum \cite{Brown:1999yd}.
We apply this strategy to the cases with staggered and Wilson fermions
on a lattice with a small aspect ratio in order to
suppress thermal excitations.
Thus we
identify the critical coupling constant that
separates the CSC phase and the normal phase,
which
exhibits many peaks as a function of the chemical potential
reflecting the discretized energy levels of quarks in a finite system.
We also obtain the form of the Cooper pair condensate at the critical point,
and investigate its momentum components and flavor structure.
Part of the results has been presented in a proceedings article \cite{Yokota:2021wwv}.

The rest of this paper is organized as follows.
In Section \ref{sec:form}
we explain the general formalism which enables us
to investigate the CSC on the lattice.
In particular, we derive the condition
for determining the critical point
from the gap equation.
In Section \ref{sec:cp} we present our numerical
results for
the critical coupling constant and the form of the Cooper pair condensate
in the case of staggered fermions
and
Wilson fermions.
Section \ref{sec:conc} is devoted to a summary and discussions.
In the appendices, we provide some details of the gap equation
and the method used to solve it.

\section{The general formalism \label{sec:form}}

In this section we derive
the gap equation, which is a self-consistent equation for the Cooper pair condensate.
The critical point is obtained from it
by assuming a continuous phase transition,
which implies that the condensate vanishes at the critical
point\footnote{Strictly speaking, phase transitions are obscured in finite systems
due to statistical fluctuations, which are suppressed by $O(1/\sqrt{V})$
for system size $V$.
These fluctuations are ignored in the mean-field approach like the one we adopt.
In fact,
one can
incorporate these fluctuations by adopting
the phenomenological Landau
prescription \cite{moretto_pairing_1972,goodman_statistical_1984}
or
the static path approximation \cite{rossignoli_effective_1994}.
Analyses based on such approaches are left for future
investigations.}.

\subsection{Derivation of the gap equation}
The gap equation is a self-consistent equation for the Cooper pair condensate,
which we derive below. For that, it is useful to work in the Nambu-Gor'kov formalism \cite{nam60,bar77,bailin_superfluidity_1981}
based on the Nambu basis
\begin{align}
	\Psi_\rho^a(N)
	=
	\begin{pmatrix}
		\psi_\rho^a(N)
		\\
		\overline{\psi}_\rho^a(N)
	\end{pmatrix}
	\, , \quad 
	\overline{\Psi}_\rho^a(N)
	=
	\begin{pmatrix}
		\overline{\psi}_\rho^a(N)
		&
		\psi_\rho^a(N)
	\end{pmatrix} 
\end{align}
and its propagator
\begin{align}
	{\bf S}^{aa'}_{\rho\rho'}(N,N')
	=
	\Braket{\Psi_\rho^a(N)\overline{\Psi}_{\rho'}^{a'}(N')}
	=
	\begin{pmatrix}
		\Braket{\psi_\rho^a(N)\overline{\psi}_{\rho'}^{a'}(N')}
		&
		\Braket{\psi_\rho^a(N)\psi_{\rho'}^{a'}(N')}
		\\
		\Braket{\overline{\psi}_\rho^a(N)\overline{\psi}_{\rho'}^{a'}(N')}
		&
		\Braket{\overline{\psi}_\rho^a(N)\psi_{\rho'}^{a'}(N')}
	\end{pmatrix} \ .
\end{align}
Here $\Braket{\ldots}$ implies
taking the quantum and thermal average and the fermion fields are represented by $\psi^{a}_{\rho}(N)$ and $\overline{\psi}^{a}_{\rho}(N)$, where $N$ and $a$
are the indices for the lattice site and color,
while $\rho$ represents the flavor and spinor indices collectively.
Assuming that the lattice translational symmetry is not spontaneously broken, the propagator in the momentum space
is given by
\begin{align}
	&\tilde{\bf S}^{aa'}_{\rho\rho'}(p,p')
	=
	\Braket{\tilde{\Psi}_\rho^a(p)\tilde{\overline{\Psi}}_{\rho'}^{a'}(p')}
	=
	\delta_{p+p'}\tilde{\bf S}^{aa'}_{\rho\rho'}(p)
	\notag
	\\
	&
	=
	\delta_{p+p'}
	\begin{pmatrix}
		\tilde{S}_{11,\rho\rho'}^{aa'}(p)
		&
		\tilde{S}_{12,\rho\rho'}^{aa'}(p)
		\\
		\tilde{S}_{21,\rho\rho'}^{aa'}(p)
		&
		\tilde{S}_{22,\rho\rho'}^{aa'}(p)
	\end{pmatrix}
	=
	\delta_{p+p'}
	\begin{pmatrix}
		\Braket{\tilde{\psi}_\rho^a(p)\tilde{\overline{\psi}}_{\rho'}^{a'}(-p)}
		&
		\Braket{\tilde{\psi}_\rho^a(p)\tilde{\psi}_{\rho'}^{a'}(-p)}
		\\
		\Braket{\tilde{\overline{\psi}}_\rho^a(p)\tilde{\overline{\psi}}_{\rho'}^{a'}(-p)}
		&
		\Braket{\tilde{\overline{\psi}}_\rho^a(p)\tilde{\psi}_{\rho'}^{a'}(-p)}
	\end{pmatrix}  \ ,
\end{align}
where $p$ and $p'$ represent the lattice momenta, and the Fourier components are
defined as
\begin{align}
	\tilde{f}(p)
	=&
	\sum_{N}e^{-ip\cdot N}f(N)
	\quad \quad
	\text{for~}
	f=\psi_\rho^a,\ \overline{\psi}_\rho^a,\ \Psi_\rho^a,\
         \overline{\Psi}_\rho^a \ .
\end{align}
The off-diagonal parts of the propagator correspond to the Cooper pair condensate.
One of the relations satisfied by the propagator is the Dyson equation
\begin{align}
	\label{eq:dyson}
	\tilde{\bf S}^{-1,aa'}_{\rho\rho'}(p)
	= \tilde{\bf D}_{\rho\rho'}^{aa'}(p) + \tilde{\bf \Sigma}_{\rho\rho'}^{aa'}(p) \ . 
\end{align}
On the right-hand side, the first term is given by 
\begin{align}
	\tilde{\bf D}_{\rho\rho'}^{aa'}(p)
	=
	\begin{pmatrix}
		\tilde{D}_{11, \rho\rho'}^{aa'}(p) & 0
		\\
		0 & \tilde{D}_{22, \rho\rho'}^{aa'}(p)
	\end{pmatrix} 
	=
	\begin{pmatrix}
		\tilde{D}_{\rho\rho'}^{aa'}(p) & 0
		\\
		0 & -\tilde{D}_{\rho'\rho}^{a'a}(-p)
	\end{pmatrix}
        \ ,
\end{align}
where $\tilde{D}_{\rho\rho'}^{aa'}(p)=\delta_{aa'}\tilde{D}_{\rho\rho'}(p)$
represents the inverse of the free-quark propagator,
and the second term represents the self-energy
\begin{align}
	\tilde{\bf \Sigma}_{\rho\rho'}^{aa'}(p)
	=
	\begin{pmatrix}
		\tilde{\Sigma}_{11,\rho\rho'}^{aa'}(p)
		&
		\tilde{\Sigma}_{12,\rho\rho'}^{aa'}(p)
		\\
		\tilde{\Sigma}_{21,\rho\rho'}^{aa'}(p)
		&
		\tilde{\Sigma}_{22,\rho\rho'}^{aa'}(p)
	\end{pmatrix} \ ,
\end{align}
whose diagonal and off-diagonal parts are associated with the chiral condensate
and the superconducting gap, respectively.

\begin{figure}[!t]
  \begin{center}
	\begin{alignat*}{3}
	\tilde{\bf \Sigma}
	=
	&
	\parbox[c]{6em}{\includegraphics[width=6em]{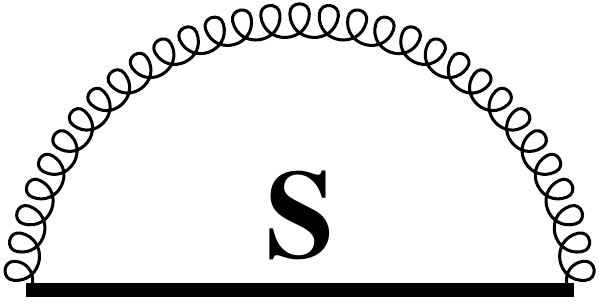}}
	+
	O(g^4)
	=
	\begin{pmatrix}
		\parbox[c]{6em}{\includegraphics[width=6em]{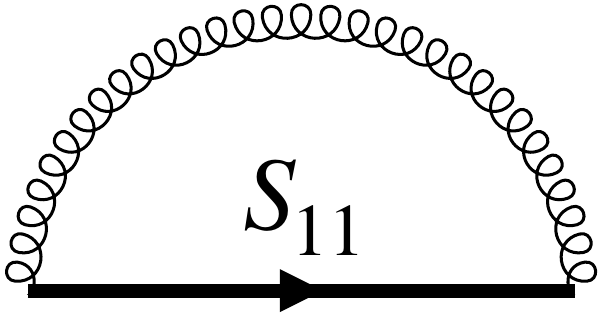}}
		&
		\parbox[c]{6em}{\includegraphics[width=6em]{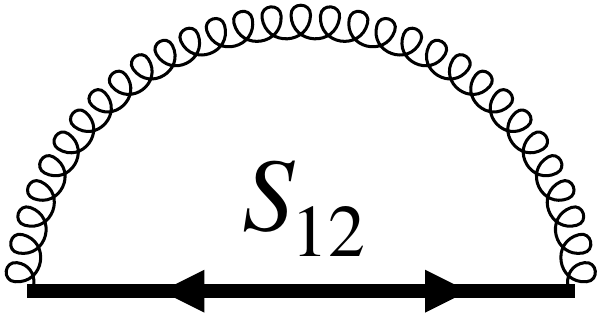}}
		\\
		\parbox[c]{6em}{\includegraphics[width=6em]{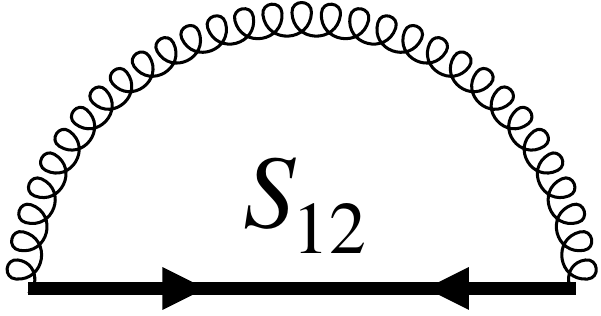}}
		&
		\parbox[c]{6em}{\includegraphics[width=6em]{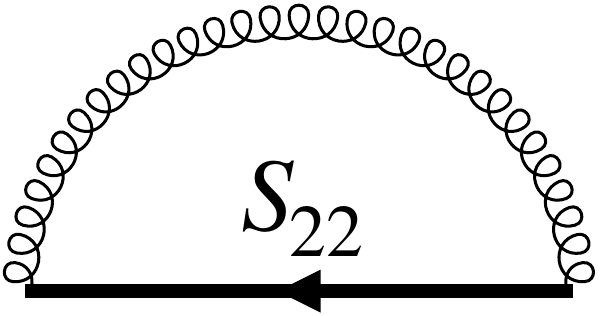}}
	\end{pmatrix}
	+
	O(g^4)
	\end{alignat*}
        \caption{Diagrammatic representation of the one-loop contribution
          to the self-energy $\tilde{\bf \Sigma}$, where $g$ represents the bare
          gauge coupling constant and the curly line represents the gluon
          propagator.}
    \label{fig:loop}
  \end{center}
\end{figure}

Since $\tilde{\bf S}$ and $\tilde{\bf \Sigma}$ are the unknowns
in Eq.~\eqref{eq:dyson}, 
another relation is needed to determine them.
In the weak coupling regime, we obtain
a relation depicted in Fig.~\ref{fig:loop}
to the lowest order in the loop expansion.
This relation together with Eq.~\eqref{eq:dyson} forms the gap equation.
In particular, the off-diagonal parts determine the superconducting gap $\tilde{\Sigma}_{12(21)}$.
When the gap equation has a non-trivial solution $\tilde{\Sigma}_{12(21)}\neq 0$ with the free energy smaller than that for the trivial solution $\tilde{\Sigma}_{12(21)}=0$, the superconducting phase is implied.

\subsection{Linearizing the gap equation}

\begin{figure}[!t]
  \begin{center}
	\begin{alignat*}{3}
	\mathrm{(a)}
	\qquad
	&
	\tilde{\Sigma}_{12,\rho\rho'}^{aa'}(p)
	&=
	\parbox[c]{12em}{\includegraphics[width=12em]{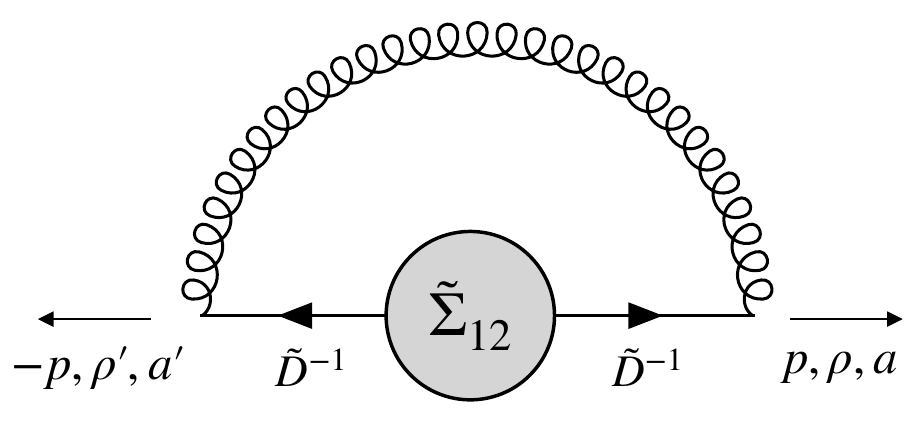}}
	\\
	\mathrm{(b)}
	\qquad
	&
	\frac{1}{\beta} \mathcal{M}_{(p\rho\rho')(q\sigma\sigma')}
	&=
	\parbox[c]{12em}{\includegraphics[width=12em]{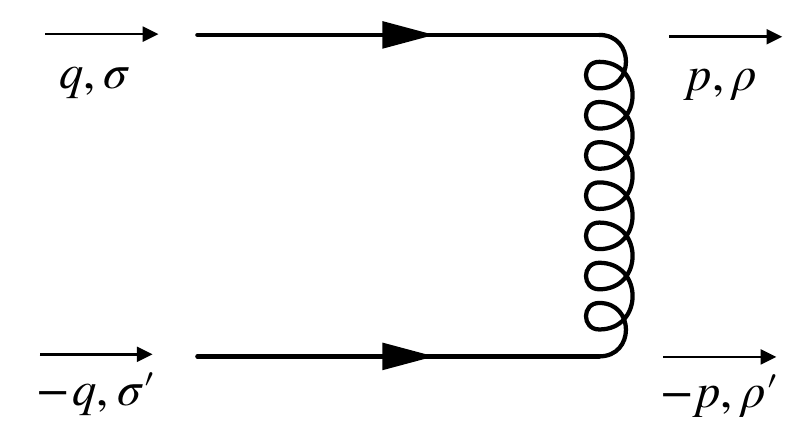}}
	\end{alignat*}
    \caption{Diagrammatic representation of the gap equation for (a) the off-diagonal self-energy
    $\tilde{\Sigma}_{12,\rho\rho'}^{aa'}(p)$
      and (b) the $\beta$-independent matrix $\mathcal{M}_{(p\rho\rho')(q\sigma\sigma')}$
      with $\beta = 2 N_c / g^2$.
      The colors of the incoming and outgoing quark pairs are anti-symmetrized.
      The curly and solid lines
      represent the gluon propagator and the free fermion propagator $\tilde{D}^{-1,aa'}_{\rho\rho'}(p)$,
      respectively.}
    \label{fig:sm_diag}
  \end{center}
\end{figure}
\begin{figure}[b]
  \begin{center}
	\begin{align*}
	\parbox[c]{5em}{\includegraphics[width=5em]{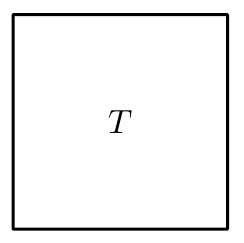}}
	&=	
	\parbox[c]{1.5em}{\includegraphics[width=1.5em]{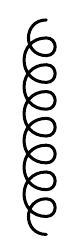}}
	+
	\parbox[c]{6em}{\includegraphics[width=6em]{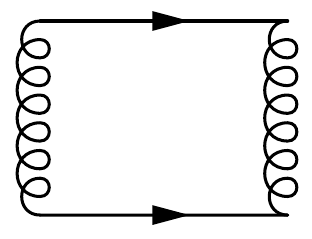}}
	+
	\cdots
	=
	\parbox[c]{1.5em}{\includegraphics[width=1.5em]{fig_gluon_vt}}
	\cdot
	\left(
	1
	-
	\frac{1}{\beta}
	\mathcal{M}
	\right)^{-1}
	\end{align*}
    \caption{Diagrammatic representation of the T-matrix.}
    \label{fig:Tmatrix}
  \end{center}
\end{figure}
Let us focus on the vicinity of the critical point,
where the non-trivial solution for $\tilde{\Sigma}_{12(21)}$ is close to zero.
We
assume that
the chiral condensate 
$\tilde{\Sigma}_{11(22)}$
is zero
and ignore higher-order corrections to $\tilde{S}_{12(21)}$.
Then, from Eq.~\eqref{eq:dyson}, we obtain
\begin{align}
	\label{eq:prop12_cp}
	\tilde{S}_{12,\rho\rho'}^{aa'}(p)
	=&
	\left[-
	\tilde{D}_{11}^{-1}(p)
	\tilde{\Sigma}_{12}(p)
	\tilde{D}_{22}^{-1}(p)
	\right]_{\rho\rho'}^{aa'}
	+
	\mathcal{O}\left(\tilde{\Sigma}_{12(21)}(p)^2\right) \  ,
	\\
	\label{eq:prop21_cp}
	\tilde{S}_{21,\rho\rho'}^{aa'}(p)
	=&
	\left[-
	\tilde{D}_{22}^{-1}(p)
	\tilde{\Sigma}_{21}(p)
	\tilde{D}_{11}^{-1}(p)
	\right]_{\rho\rho'}^{aa'}
	+
	\mathcal{O}\left(\tilde{\Sigma}_{12(21)}(p)^2\right) \ ,
\end{align}
which makes the gap equation linear in $\tilde{\Sigma}_{12(21)}$.
For instance, the equation for $\tilde{\Sigma}_{12}$ is given diagrammatically
in Fig.~\ref{fig:sm_diag}(a).
Let us note
that the vertex is proportional to $gT^{B}_{aa'}$,
where $B$ represents the color charge of the gluon and $T^{B}_{aa'}$ represents
a generator of $\mathrm{SU}(N_{\rm c})$ with $N_{\rm c}$ being the number of colors.
By using the identity
\begin{align}
	\label{eq:Tsum}
    g^2\sum_{B}T^B_{ab}T^B_{a'b'}
    =
    \frac{N_{\rm c}-1}{2\beta}
    (
    \delta_{ab}\delta_{a'b'}
    +
    \delta_{ab'}\delta_{a'b}
    )
    -
    \frac{N_{\rm c} + 1}{2\beta}
    (
    \delta_{ab}\delta_{a'b'}
    -
    \delta_{ab'}\delta_{a'b}
    ) \ ,
\end{align}
where $\beta=2N_{\rm c}/g^2$, we can
decompose
the gap equation into those
for the color-symmetric part $\tilde{\Sigma}_{12(p\rho\rho')}^{(+)aa'}$ and
the color-antisymmetric part $\tilde{\Sigma}_{12(p\rho\rho')}^{(-)aa'}$ as
\begin{align}
  	\tilde{\Sigma}^{(\pm)aa'}_{12,(p \rho\rho')}
	=
	\frac{\tilde{\Sigma}^{aa'}_{12,\rho\rho'}(p)
	\pm
	\tilde{\Sigma}^{a'a}_{12,\rho\rho'}(p)
	}{2} \ .
\end{align}
The color-symmetric and antisymmetric terms in Eq.~\eqref{eq:Tsum} have different signs,
which reflects the fact that the interaction is repulsive (attractive)
in the color-symmetric (antisymmetric) channel.
Since the Cooper instability occurs in the attractive channels,
we will concentrate on the off-diagonal self-energy in the color
anti-symmetric channel $\tilde{\Sigma}_{12(p\rho\rho')}^{(-)aa'}$ from now on.
Also, we will suppress the color indices
because the equation has the same form for any choice of colors.
Thus, the gap equation becomes
\begin{align}
	\label{eq:gap_linear}
	\sum_{q\sigma\sigma'}
	\mathcal{M}_{(p\rho\rho')(q\sigma\sigma')}
	\tilde{\Sigma}^{(-)}_{12(q\sigma\sigma')}
	=
	\beta \, 
	\tilde{\Sigma}^{(-)}_{12(p\rho\rho')} \ ,
\end{align}
where the $\beta$-independent matrix $\mathcal{M}_{(p\rho\rho')(q\sigma\sigma')}$ is defined
in Fig.~\ref{fig:sm_diag}(b).
The explicit forms of $\mathcal{M}_{(p\rho\rho')(q\sigma\sigma')}$ for
staggered and Wilson fermions
are given in Appendix~\ref{sec:m_form}.
The largest eigenvalue $\lambda_1$ of $\mathcal{M}$ is identified
as the critical value
of $\beta$
\begin{align}
	\label{eq:critical_cond}
        	\beta_{\rm c}=\lambda_1[\mathcal{M}]
\end{align}
since no condensate occurs above this value; i.e.,
the system is in a normal phase at weaker coupling.
In order to obtain the
largest eigenvalue,
we use the power iteration method
as explained in Appendix \ref{sec:power-iter}.
From the eigenvector
corresponding to
the largest eigenvalue $\beta_{\rm c}$,
we obtain the form of the Cooper pair condensate at the critical point.
Note also that 
the use of $\tilde{\Sigma}_{21}$ instead of $\tilde{\Sigma}_{12}$
leads to the same condition.

The condition~\eqref{eq:critical_cond} can be
regarded as a generalization of
the Thouless criterion \cite{tho60}, which is given by the divergence of
the T-matrix
for specific types of interaction.
Indeed, the T-matrix given diagrammatically in Fig.~\ref{fig:Tmatrix} implies
that Eq.~\eqref{eq:gap_linear}
is equivalent
to the divergence of the T-matrix.

\section{Results for the critical point and
  the Cooper pair condensate \label{sec:cp}}

In this section we use
the general formalism in the previous section
to determine the parameter region for the CSC
and the form of the Cooper pair condensate
at the critical point.
Our results for the critical point include
large values of $\beta$, which
correspond to a small physical size of the system
compared to $\Lambda_{\rm QCD}^{-1}$.
In that case,
our weak coupling analysis is justifiable
and provides
an excellent testing ground for
non-perturbative approaches such as the complex Langevin method.
In Sections \ref{sec:stag_beta} and \ref{sec:stag_coop}
we discuss the case of four-flavor staggered fermions,
which has the advantage of maintaining some part of chiral symmetry explicitly.
In Section \ref{sec:res_wil}
we show our results in the case of Wilson fermions,
which has the advantage of applicability to any number of flavors.

\subsection{The critical point for staggered fermions \label{sec:stag_beta}}
\begin{figure}[t]
  \begin{center}
  	\includegraphics[width=0.7\columnwidth]{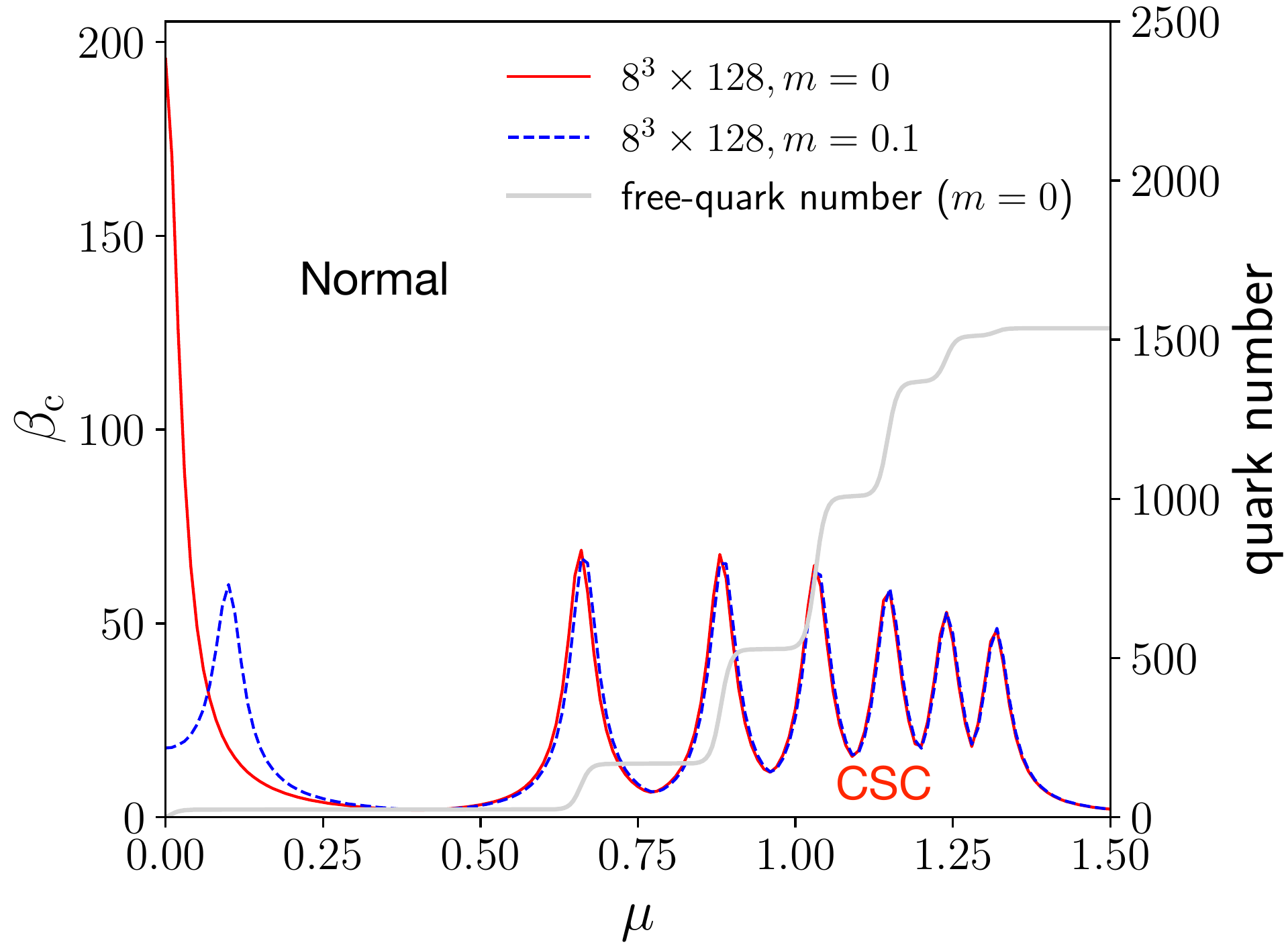}
    \caption{
      The phase diagram in the $\mu$-$\beta$ plane.
      The region above the critical coupling $\beta_c$ corresponds to
      the normal phase, whereas the region below $\beta_c$ corresponds to 
      the CSC phase. The results for staggered fermions on an $8^3\times 128$
      lattice with
      $m=0$ (red solid line) and $m=0.1$ (blue dashed line) are shown.
      The solid gray line represents the quark number $N_{\rm q}$ for
      free quarks with $m=0$.}
    \label{fig:l8t128}
  \end{center}
\end{figure}

Here we identify the boundary of the normal and CSC phases characterized by $\beta_c$
as a function of the quark chemical potential $\mu$
on lattices $L_s^3 \times L_t = 4^3 \times 64$, $8^3 \times 128$ and $16^3 \times 256$.
The chosen aspect ratio $L_{s}/L_{t} = 1/16$
is small enough to suppress the thermal fluctuations
that may destroy the CSC.

Figure \ref{fig:l8t128} shows the phase diagram in the $\mu$-$\beta$ plane
for staggered fermions
on an $8^3\times 128$ lattice
with quark mass $m=0$ and $0.1$ in lattice units.
The existence of the CSC is suggested below $\beta_c$ although
it may be affected by higher order corrections in the small $\beta$ region.
The obtained $\beta_c$ has many peaks as a function of $\mu$
similarly to the results for the NJL model
in a finite box \cite{amo02}. Each peak corresponds to the enhancement of
the energy gap that occurs when
$\mu$
is close to an energy level of the free quark
\begin{align}
	E({\bm p})=\sinh^{-1}\sqrt{\sum_{i=1}^{3}\sin^{2} p_{i}+m^2} \ ,
\end{align}
where the momentum in the first Brillouin zone is given by
\begin{align}
	\label{eq:bz_staggered}
	\lbrace
	{\bm p}|{\bm p}\in {\rm BZ_{\rm s}}
	\rbrace
	=&
	\left\lbrace
	\left.
	\left(
	\frac{2\pi n_1}{L_{s}},
	\frac{2\pi n_2}{L_{s}},
	\frac{2\pi n_3}{L_{s}}
	\right)
	\right|
	-\frac{L_{s}}{4}
	\leq n_{i}
	<\frac{L_{s}}{4},
	n_{i}\in \mathbb{Z}
	\right\rbrace \ .
\end{align}
The size of the momentum space is half ($L_{s}/2$) of the spatial lattice size 
since staggered fermions correspond to Dirac fermions on a coarser lattice with
twice as large lattice spacing as the original one.
Note that the peak at $\mu=0$ that appears for $m = 0$
shifts for finite $m$ to the location corresponding
to the lowest energy level $E(\bm{0})=\sinh^{-1} m$.
This peak is considered to be a finite-size artifact
since it is caused by the condensate of
quark-quark and antiquark-antiquark pairs with zero momentum,
which vanishes in the thermodynamic limit $L_s \rightarrow \infty$.

In Fig.~\ref{fig:l8t128} we also plot the quark number for free quarks given by 
\begin{align}
	N_{\rm q}
	=
	N_{\rm sp}N_{\rm c}N_{\rm f}
	\sum_{{\bm p}\in {\rm BZ}_{\rm s}}
	\left[
	n_{\rm F}(E({\bm p})-\mu)
	-
	n_{\rm F}(E({\bm p})+\mu)
	\right] \ ,
\end{align}
where $N_{\rm sp}=2$, $N_{\rm c}=3$ and $N_{\rm f}=4$ represent
the spin, color, and flavor degrees of freedom, respectively,
and $n_{\rm F}(x)=[\exp(x/T)+1]^{-1}$ represents the Fermi distribution function
at temperature $T$. 
As $\mu$ increases, the Fermi sphere becomes larger and includes more high-momentum modes,
which leads to the stepwise
increase
of
$N_{\rm q}$ \cite{matsuoka_thermal_1984};
\emph{i.e.},
the quark number $N_{\rm q}$ jumps when $\mu$ reaches
$\mu=E(\bm{p})$, a discrete energy level of quarks.
The critical $\beta_c$ has a peak at $\mu$ corresponding to the jump of $N_{\rm q}$.
This is
consistent with the picture that
the condensate is mainly caused by the scattering of fermions near the Fermi surface.

\begin{figure}[!t]
  \begin{center}
  	\includegraphics[width=\columnwidth]{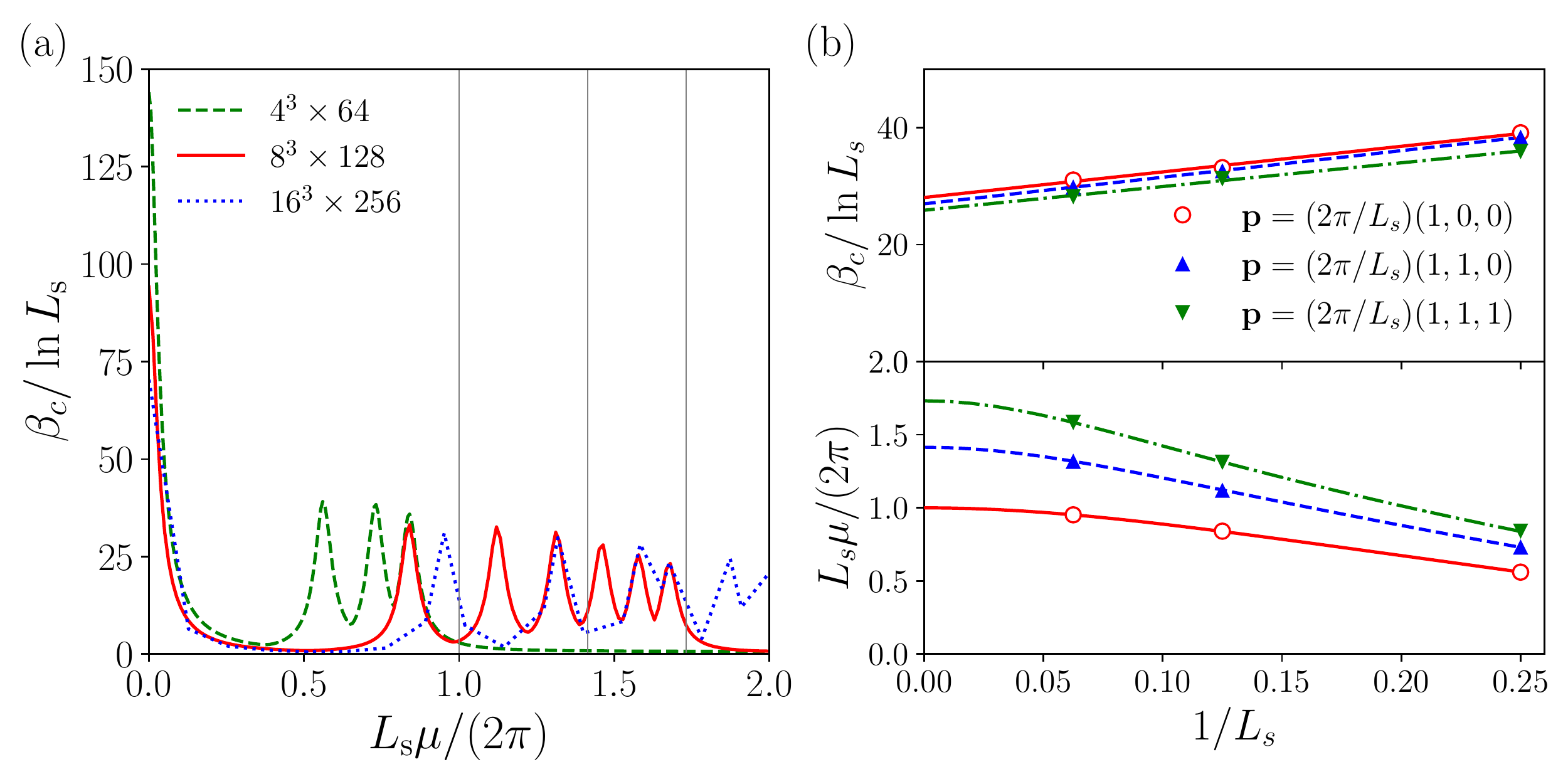}
        \caption{(a) The critical coupling $\beta_{c}$ is plotted against
        the quark chemical potential $\mu$
        for staggered fermions with quark mass $m=0$ on various lattices
        $L_s=4,8,16$ with a fixed aspect ratio $L_s / L_t = 1 / 16$.
        Appropriate normalization is used to reveal the expected scaling behavior
        at large $L_s$.
        The gray vertical lines indicate the expected positions of the peaks
        in the $L_{s} \to \infty$ limit.
          (b) The height $\beta_{c} / \ln L_{s}$ and the position $L_{s}\mu /(2\pi)$
          of the peaks
          are plotted against $1 / L_{s}$
          for momentum modes ${\bm p}=(2\pi/L_s)(1,0,0),\, (2\pi/L_s)(1,1,0),\,(2\pi/L_s)(1,1,1)$.
          The lines in the plot below
          represent $\mu=E(\boldsymbol{p})$.
          }
    \label{fig:beta_scaling}
  \end{center}
\end{figure}

Next, let us discuss
the lattice-size dependence of $\beta_c$.
By using the dimensional analysis and the $\beta$ function at the one-loop level,
we expect a scaling behavior
\begin{align}
	\label{eq:b_scaling}
	\beta_c 
	\sim 
	f\left(\hat{L}\hat{\mu}, \hat{L}\hat{T}\right)\ln\left(\frac{1}{a\Lambda_{\rm QCD}}\right)
	=
	f\left(L_s \mu, \frac{L_s}{L_t}\right)\ln\left(\frac{L_s}{\hat{L}\Lambda_{\rm QCD}}\right) \ ,
\end{align}
where $a$ is the lattice spacing, $f$ is a dimensionless function,
while $\hat{L}=aL_s$, $\hat{\mu}=\mu/a$ and $\hat{T}=1/(aL_t)$ are
the dimensionful spatial extent, the quark chemical potential, and the temperature, respectively.
Eq.~\eqref{eq:b_scaling} suggests
\begin{align}
	\label{eq:betals}
	\beta_{c} \propto \ln L_{s}
\end{align}
with the dimensionful quantities $\hat{\mu}$, $\hat{T}$ and $\hat{L}$ fixed,
or equivalently, $L_s \mu$, $L_s/L_t$ and $\hat{L}$ fixed.
Figure \ref{fig:beta_scaling}(a) shows the lattice-size dependence of $\beta_c$
for a fixed aspect ratio $L_{s}/L_{t}=1/16$.
In Fig.~\ref{fig:beta_scaling}(b)
we show the $L_s$ dependence of the height and the position of the peaks
corresponding to the momentum
${\bm p}=(2\pi/L_s)(1,0,0),\, (2\pi/L_s)(1,1,0),\,(2\pi/L_s)(1,1,1)$.
For all momenta, $\beta_c/\ln L_{s}$ depends linearly on $1/L_s$ and
converges to a finite value as $L_s\to\infty$, which suggests
that the height of each peak scales as $\ln L_{s}$ for large $L_s$. 
The peak positions agree with $\mu=E({\bm p})$ and
converge to $L_{s}\mu/(2\pi)=\sqrt{n_1^2+n_2^2+n_3^2}$ ($n_{1,2,3}\in \mathbb{Z}$)
as $L_s \rightarrow \infty$.
In fact, we find that the peak height $\beta_c$
at $\mu=0$ is almost independent of $L_s$ and it
does not follow
the scaling \eqref{eq:betals} unlike the other peaks\footnote{See the decrease of the peak height at $\mu=0$ with increasing $L_s$.},
which is consistent with our aforementioned interpretation that the peak at $\mu=0$
is merely a finite-size artifact.

\subsection{The Cooper pair condensate for staggered fermions \label{sec:stag_coop}}

\begin{figure}[b]
  \begin{center}
  	\includegraphics[width=0.7\columnwidth]{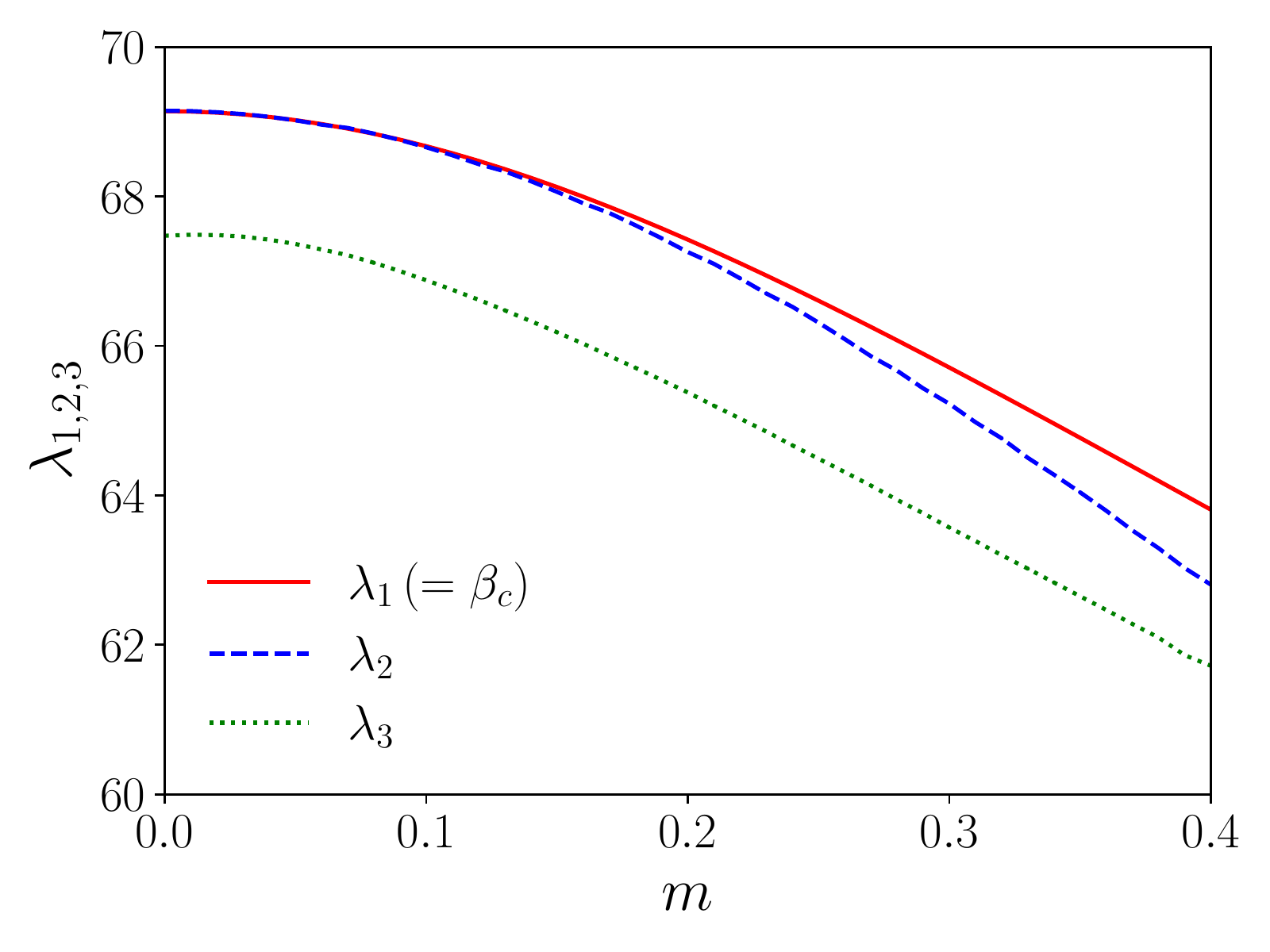}
        \caption{The three largest eigenvalues $\lambda_{1,2,3}$
          of $\mathcal{M}$
       for staggered fermions on an $8^3\times 128$ lattice
                    are plotted against the quark mass $m$
          at $\mu = E(|{\bm p}| = 2 \pi / L_s)$.
    }
    \label{fig:beta_mqdep}
  \end{center}
\end{figure}

\begin{table}[htb]
  \centering
  \caption{The irreducible representations of the Euclidean Lorentz group
    for the Cooper pairs.
    The condensate and the number of components $N_{\rm comp}$ that correspond to each representation are also shown. We use the Euclidean Dirac gamma matrices
    $\gamma_\dnu$, $\gamma_5=\gamma_1\gamma_2\gamma_3\gamma_4$,
    $\sigma_{\dnu\dnu'}=(i/2)[\gamma_\dnu,\gamma_{\dnu'}]$ and the charge conjugation
    operator $C=\gamma_2\gamma_4$.
    Note that $N_{\rm comp}=3$ for $\tilde{S}_{\rm t(pt),\dnu\dnu'}^{fg}(p)$ is obtained from the constraints 
    $\tilde{S}_{\rm t(pt),\dnu\dnu'}^{fg}(p)=-\tilde{S}_{\rm t(pt),\dnu'\dnu}^{fg}(p)$ and 
    $\tilde{S}_{\rm t(pt),\dnu_1\dnu_2}^{fg}(p)
    =
    (i/2)\epsilon_{\dnu_1\dnu_2\dnu_3\dnu_4}\tilde{S}_{\rm t(pt),\dnu_3\dnu_4}^{fg}(p)$.
    }
  \label{tab:lorentz}
{\renewcommand\arraystretch{1.8}
    \begin{tabular}{ccc}
	  \hline
      representation
      &
      condensate
      &
      $N_{\rm comp}$
      \\
      \hline
      \hline 
      \multirow{2}{*}{scalar}
      &
      $
      \tilde{S}_{\rm s}^{fg}(p)
      =
      \epsilon_{ab3}\Braket{\tilde{\psi}_f^{a}(p)\Gamma^{\rm s} \tilde{\psi}_g^{b}(-p)}$
      &
      \multirow{2}{*}{1}
      \\
      &
      $\Gamma^{\rm s}=\gamma_5 C$
      \\
      \hline
      \multirow{2}{*}{pseudo-scalar}
      &
      $
      \tilde{S}_{\rm ps}^{fg}(p)
      =
      \epsilon_{ab3}
      \Braket{\tilde{\psi}_f^{a}(p)\Gamma^{\rm ps} \tilde{\psi}_g^{b}(-p)}
      $
      &
      \multirow{2}{*}{1}
      \\
      &
      $\Gamma^{\rm ps}=C$
      \\
      \hline 
      \multirow{2}{*}{vector}
      &
      $
      \tilde{S}_{\rm v,\dnu}^{fg}(p)
      =
      \epsilon_{ab3}
      \Braket{\tilde{\psi}_f^{a}(p)\Gamma^{\rm v,\dnu} \tilde{\psi}_g^{b}(-p)}
      $
      &
      \multirow{2}{*}{4}
      \\
      &
      $\Gamma^{\rm v,\dnu}=C\gamma_5\gamma_\dnu$
      \\
      \hline
      \multirow{2}{*}{pseudo-vector}
      &
      $
      \tilde{S}_{\rm pv,\dnu}^{fg}(p)
      =
      \epsilon_{ab3}
      \Braket{\tilde{\psi}_f^{a}(p)\Gamma^{\rm pv,\dnu} \tilde{\psi}_g^{b}(-p)}
      $
      &
      \multirow{2}{*}{4}
      \\
      &
      $\Gamma^{\rm pv,\dnu}=C\gamma_\dnu$
      \\
      \hline
      \multirow{2}{*}{\parbox{4cm}{\centering self-dual antisymmetric tensor}}
      &
      $
      \tilde{S}_{\rm t,\dnu\dnu'}^{fg}(p)
      =
      \epsilon_{ab3}
      \Braket{\tilde{\psi}_f^{a}(p)\Gamma^{\rm t,\dnu\dnu'} \tilde{\psi}_g^{b}(-p)}
      $
      &
      \multirow{2}{*}{3}
      \\
      &
      $\Gamma^{\rm t,\dnu\dnu'}=C\gamma_{5}\sigma_{\dnu\dnu'}$
      \\
      \hline
      \multirow{2}{*}{\parbox{4cm}{\centering pseudo-self-dual antisymmetric tensor}}
      &
      $
      \tilde{S}_{\rm pt,\dnu\dnu'}^{fg}(p)
      =
      \epsilon_{ab3}
      \Braket{\tilde{\psi}_f^{a}(p)\Gamma^{\rm pt,\dnu\dnu'} \tilde{\psi}_g^{b}(-p)}
      $
      &
      \multirow{2}{*}{3}
      \\
      &
      $\Gamma^{\rm pt,\dnu\dnu'}=C\sigma_{\dnu\dnu'}$
      \\
      \hline
    \end{tabular}
}
\end{table}

The Cooper pair condensate at the critical point can be obtained from
the eigenvector corresponding to the largest eigenvalue
$\beta_c$ of $\mathcal{M}$ through Eq.~\eqref{eq:prop12_cp} up to an overall factor.
We define the Cooper pair condensate in the momentum space as
\begin{align}
	\label{eq:sdef}
	\tilde{S}_{\alpha\beta}^{fg}(p)
	\equiv
	\sum_{a,b}
	\epsilon_{abc}
	\tilde{S}_{12,(f\alpha)(g\beta)}^{ab}(p)
        =
	\sum_{a,b}
	\epsilon_{abc}
	\Big\langle
	\tilde{\psi}_{f \alpha}^{a}(p)
	\tilde{\psi}_{g \beta}^{b}(-p)
	\Big\rangle   \ , 
\end{align}
where $\tilde{\psi}_{f \alpha}^{a}(p)$ is the four-flavor Dirac fermion field with the 4d momentum $p=(\boldsymbol{p},p_4)$ constructed from the staggered fermion field with $a$, $f$ and $\alpha$ being the color, flavor and spinor indices, respectively; see Eq.~\eqref{eq:dirac_staggered}. We fix the color index to $c=3$ on the right-hand side of Eq.~\eqref{eq:sdef} since the following results do not depend on this choice.
We have confirmed that
Eq.~\eqref{eq:sdef} has large values when $\boldsymbol{p}$
satisfies $\mu \approx E(\boldsymbol{p})$ and $p_4$ is given
by the lowest Matsubara frequencies $p_4= \pm \pi/L_t$ \cite{Yokota:2021wwv},
which is consistent with the fact that
the condensate is formed by quarks with momenta near the Fermi surface.

Since the Cooper pair is a product of two Dirac spinors,
it can be
decomposed into irreducible representations of the Euclidean Lorentz group as
\begin{align}
	\tilde{S}^{fg}_{\alpha\beta}(p)
	= \,
	&
	\frac{1}{4}
	\Gamma_{\alpha\beta}^{\rm s}
	\tilde{S}^{fg}_{\rm s}(p)
	+
	\frac{1}{4}
	\Gamma_{\alpha\beta}^{\rm ps}
	\tilde{S}^{fg}_{\rm ps}(p)
	+
	\frac{1}{4}
	\sum_{\dnu}
	\Gamma_{\alpha\beta}^{\rm v,\dnu}
	\tilde{S}^{fg}_{\rm v,\dnu}(p)
	+
	\frac{1}{4}
	\sum_{\dnu}
	\Gamma_{\alpha\beta}^{\rm pv,\dnu}
	\tilde{S}^{fg}_{\rm pv,\dnu}(p)
	\notag
	\\
	&
	+
	\frac{1}{8}
	\sum_{\dnu>\dnu'}
	\Gamma_{\alpha\beta}^{\rm t,\dnu\dnu'}
	\tilde{S}^{fg}_{\rm t,\dnu\dnu'}(p)
	+
	\frac{1}{8}
	\sum_{\dnu>\dnu'}
	\Gamma_{\alpha\beta}^{\rm pt,\dnu\dnu'}
	\tilde{S}^{fg}_{\rm pt,\dnu\dnu'}(p) \ ,
\end{align}
where the quantities on the right-hand side are defined in Table \ref{tab:lorentz}. Note that $\Gamma_{\alpha\beta}^{\rm s(ps)}$ and $\Gamma_{\alpha\beta}^{\rm t(pt),\nu\nu'}$ are anti-symmetric with respect to the exchange of $\alpha$ and $\beta$, while $\Gamma_{\alpha\beta}^{\rm v(pv),\nu}$ are symmetric.
Strictly speaking, the Euclidean Lorentz symmetry is broken to the discrete rotational group on the lattice.
However, the violation is expected to be small for $\beta_c$ at $\mu$ corresponding to the peaks (See the values of $\beta_c$
in Fig.~\ref{fig:l8t128}.).
The Dirac gamma matrices are defined, for instance, by
\begin{align}
	\gamma_{i}
	=
	\begin{pmatrix}
	0 & i\sigma^{i}
	\\
	-i\sigma^{i} & 0
	\end{pmatrix} \ ,  \quad
	\gamma_{4}
	=
	\begin{pmatrix}
	I_2 & 0
	\\
	0 & -I_2
	\end{pmatrix}
\end{align}
with the Pauli matrices $\sigma^i$ and the $2\times 2$ identity matrix $I_2$.

To investigate the components of the Cooper pair condensate, we define
\begin{align}
	\label{eq:Rs}
	R_{\rm s(ps)}
	=& \,
	A\sum_{p,f,g}
	\left|
	\tilde{S}^{fg}_{\rm s(ps)}(p)
	\right|^2,
	\\
	\label{eq:Rv}
	R_{\rm v(pv)}
	=& \,
	A\sum_{p,f,g, \dnu}\left|\tilde{S}^{fg}_{\rm v(pv),\dnu}(p)\right|^2,
	\\
	\label{eq:Rt}
	R_{\rm t(pt)}
	=& \,
	\frac{A}{2}\sum_{\dnu>\dnu'}
	\sum_{p,f,g}\left|\tilde{S}^{fg}_{\rm t(pt),\dnu\dnu'}(p)\right|^2 \ ,
\end{align}
where $A$ is a normalization factor
chosen so that
$\sum_{i={\rm s},{\rm ps},{\rm v},{\rm pv},{\rm t},{\rm pt}}R_i=1$.
The results are almost the same even if we restrict the sum over the momentum
to the region $\mu \approx E(\boldsymbol{p})$ and $p_4= \pm \pi/L_t$,
where the Cooper pair condensate becomes large.

\begin{figure}[!t]
  \begin{center}
  	\includegraphics[width=\columnwidth]{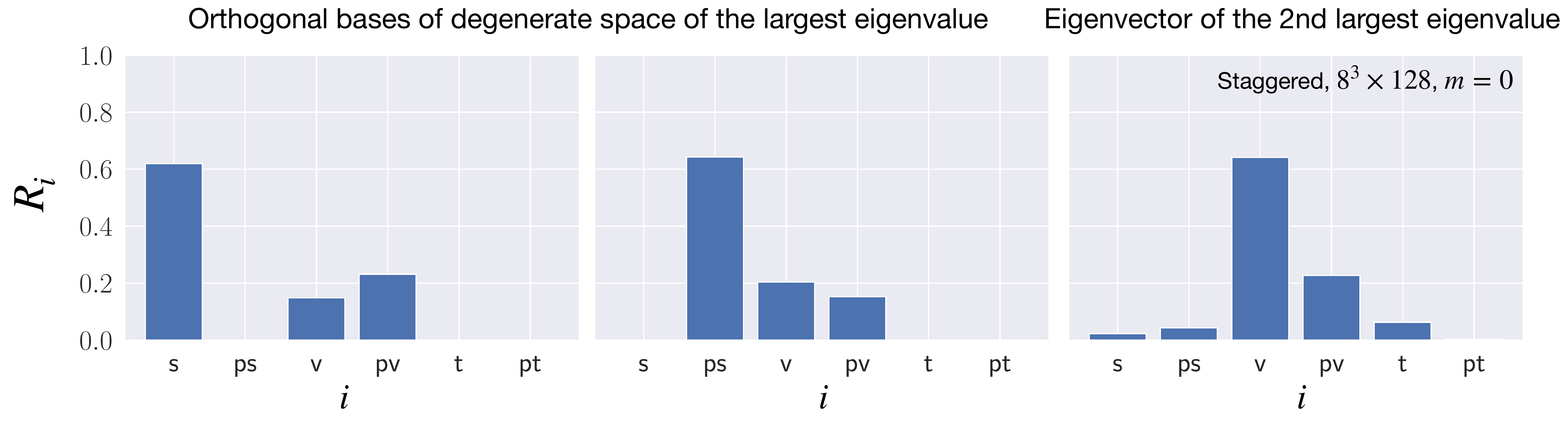}
        \caption{The components of the Cooper pair condensate
          $R_{i}$ ($i=\mathrm{s},\mathrm{ps}, \mathrm{v},\mathrm{pv}, \mathrm{t},\mathrm{pt}$)
          for staggered fermions on an $8^3\times 128$ lattice with
          $m=0$ at $\mu = E(|{\bm p}| = 2 \pi / L_s)$.
          The figures on the left and the middle show
          the results for the two orthogonal bases of
          the eigenspace of the degenerate two largest eigenvalues $\beta_c$
          as described in the text.
          The figure on the right shows
          the result for the eigenvector corresponding to
          the third largest eigenvalue.}
    \label{fig:strength_mq0}
  \end{center}
\end{figure}

In Fig.~\ref{fig:beta_mqdep} we plot the three largest eigenvalues
$\lambda_{1,2,3}$ of $\mathcal{M}$
against the quark mass.
As one approaches $m=0$,
the two largest eigenvalues come close to each other,
which
implies
the double degeneracy
at $m=0$.
In Fig.~\ref{fig:strength_mq0} we show the components
of the Cooper pair condensate
represented by $R_i$ for the eigenvectors corresponding to the eigenvalues
up to the third largest one for $m=0$. For the two largest eigenvalues,
which are degenerate as shown in Fig.~\ref{fig:beta_mqdep}, we determine
the two orthogonal bases of the eigenspace in such a way that either
the scalar or the pseudo-scalar component becomes zero.
In general, the Cooper pair condensate at the critical point
is
represented by a linear combination of the scalar and the pseudo-scalar components
with the rest of the components being small.

\begin{figure}[!t]
  \begin{center}
  	\includegraphics[width=\columnwidth]{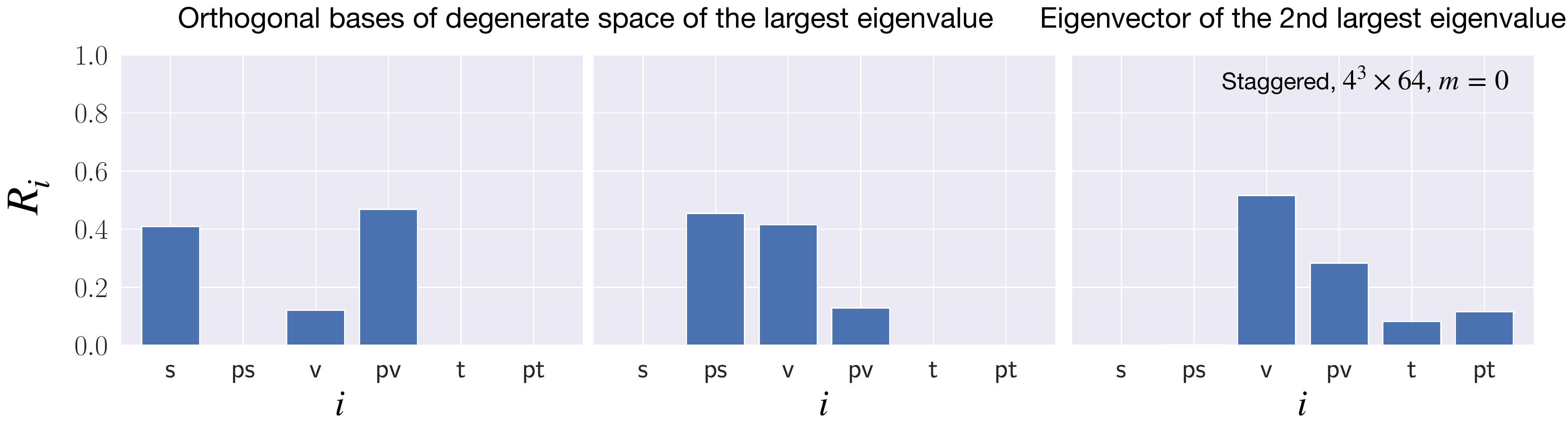}
        \caption{The same as Fig.~\ref{fig:strength_mq0} except that
          the lattice size is $4^3\times 64$.
        }
    \label{fig:strength_mq0_L4T64}
  \end{center}
\end{figure}

\begin{figure}[!t]
  \begin{center}
  	\includegraphics[width=\columnwidth]{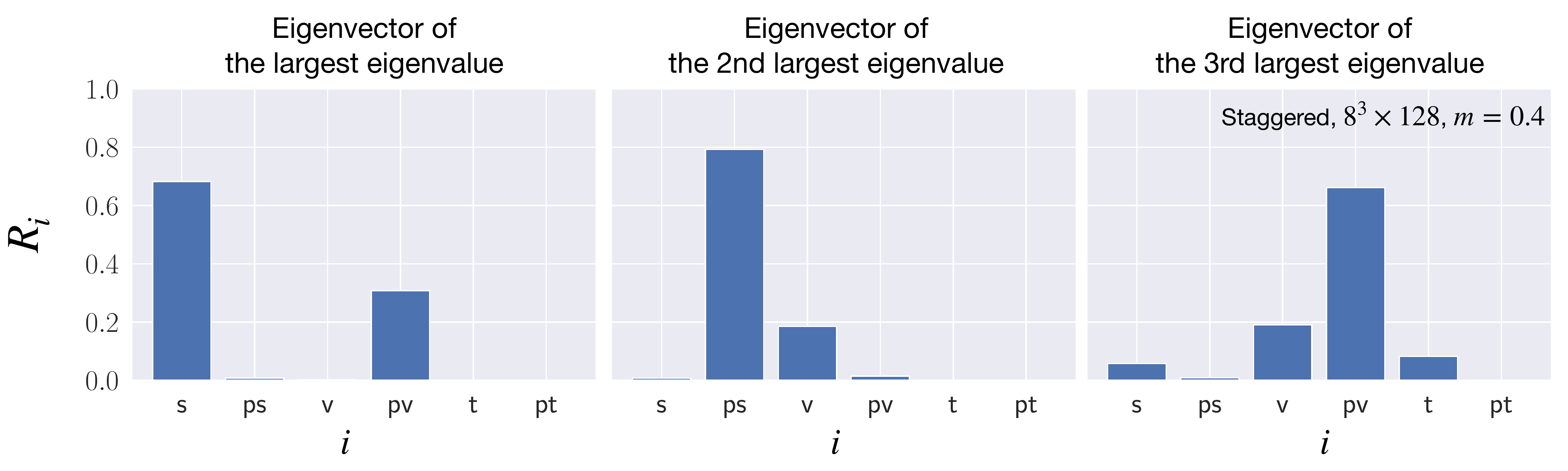}
        \caption{The components of the Cooper pair condensate
          $R_{i}$ ($i=\mathrm{s},\mathrm{ps}, \mathrm{v},\mathrm{pv}, \mathrm{t},\mathrm{pt}$)
          for the eigenvectors corresponding to the eigenvalues up to the third largest one
          for staggered fermions on an $8^3\times 128$ lattice with
          $m=0.4$ at $\mu = E(|{\bm p}| = 2 \pi / L_s)$.}
    \label{fig:strength_mq0.4}
  \end{center}
\end{figure}

In Fig.~\ref{fig:strength_mq0_L4T64} we
show the results on a $4^3\times 64$ lattice.
By comparing them with the results on an $8^3\times 128$ lattice
in Fig.~\ref{fig:strength_mq0},
we observe that the components other than the
scalar and the pseudo-scalar are suppressed
as the lattice size increases.
This suggests that the existence of
these components is due to
the breaking of the Euclidean Lorentz symmetry by the lattice discretization.
The result that the scalar or pseudo-scalar condensate is favored is consistent
with the previous work \cite{alford_qcd_1998,alford_single_2003,iwasaki_superconductivity_1995,schafer_quark_2000,schmitt_when_2002,buballa_anisotropic_2003},
which shows that pairing that breaks rotational symmetry is weaker.
Similarly, the result of $R_i$ for $m=0.4$ is shown
in Fig.~\ref{fig:strength_mq0.4}.
The degeneracy of the largest eigenvalues is lifted due to the finite mass,
and the scalar condensate is favored at the critical point
in contrast to the massless case.
This is consistent with the common wisdom that
the effect
of quark mass favors
the scalar condensate instead of the pseudo-scalar condensate.

Since the scalar and pseudo-scalar condensates
are anti-symmetric with respect to the spinor indices,
they satisfy 
\begin{align}
	\tilde{S}^{fg}_{\rm s(ps)}(p)
	=
	-\tilde{S}^{gf}_{\rm s(ps)}(-p)
        \label{Sfg-p}
\end{align}
due to the anti-commuting property of the fermion fields.
Eq.~\eqref{Sfg-p} can be rewritten as
\begin{align}
	\label{eq:Kfchange}
	\tilde{S}^{(\pm) fg}_{\rm s(ps)}(p)
	=
	\mp \tilde{S}^{(\pm) gf}_{\rm s(ps)}(p) \ ,
\end{align}
where we have defined the spatially symmetric and anti-symmetric components as
\begin{align}
	\tilde{S}^{(\pm)fg}_{\rm s(ps)}(p)
	=
	\frac{\tilde{S}^{fg}_{\rm s(ps)}(p) \pm
          \tilde{S}^{fg}_{\rm s(ps)}(-p)}{2} \ .
\end{align}
In order to determine
which component is dominant, we calculate
\begin{align}
	R^{(\pm)}_{\rm s(ps)}
	=
	A' \sum_{f,g, p}\left|\tilde{S}^{(\pm) fg}_{\rm s(ps)}(p)\right|^2 \ ,
        \label{def-R-parity}
\end{align}
where $A'$ is a normalization factor chosen
so that $R^{(+)}_{\rm s(ps)}+R^{(-)}_{\rm s(ps)}=1$.
For both the scalar and pseudo-scalar cases, we obtain
$R^{(+)}_{\rm s(ps)}\approx 0.71$
for an $8^3 \times 128$ lattice,
which implies that the spatially symmetric component $\tilde{S}^{(+)fg}_{\rm s(ps)}(p)$ is dominant.

Let us also comment (See Ref.~\cite{Yokota:2021wwv}.) that
$\tilde{S}^{(+)fg}_{\rm s(ps)}(p)$
do not depend on the direction of $\boldsymbol{p}$,
which suggests spatially isotropic s-wave superconductivity.
Note that the complex phase of $\tilde{S}^{(+)fg}_{\rm s(ps)}(p)$,
which is independent of $p$,
can take an arbitrary value,
reflecting
the spontaneous breaking of the $\mathrm{U}_{\rm B}(1)$ baryon-number symmetry
$\tilde{\psi}(p)\to e^{i\theta_{\rm B}}\tilde{\psi}(p)$ for 
$\theta_{\rm B} \in \mathbb{R}$.

Next let us focus on the flavor structure of $\tilde{S}^{(+)fg}_{\rm s(ps)}(p)$.
Since it is anti-symmetric with respect to
the exchange of $f$ and $g$ as shown in Eq.~\eqref{eq:Kfchange}, we can decompose it as
\begin{align}
	\label{eq:Kdecomp}
	\tilde{S}^{(+)fg}_{\rm s(ps)}(p)
	=& \,
	\tilde{\kappa}_{1,\rm s(ps)}(p)t_1^{fg}
	+\tilde{\kappa}_{3,\rm s(ps)}(p)t_3^{fg}
	+\tilde{\kappa}_{13,\rm s(ps)}(p)\omega_{13}^{fg}
	+\tilde{\kappa}_{24,\rm s(ps)}(p)\omega_{24}^{fg}
	\notag
	\\
	&
	+\tilde{\kappa}_{25,\rm s(ps)}(p)(t_2 t_5)^{fg}
	+\tilde{\kappa}_{45,\rm s(ps)}(p)(t_4 t_5)^{fg},
\end{align}
where $t_1$, $t_3$, $\omega_{13}$, $\omega_{24}$, $t_2 t_5$ and $t_4 t_5$ are
linearly independent anti-symmetric matrices
defined through $t_{\mu}={}^{\rm t}\gamma_{\mu}$,
$\omega_{\mu\nu}=(i/2)[t_{\mu},t_{\nu}]$,
$t_5=t_1 t_2 t_3 t_4$.
We calculate the coefficients $\tilde{\kappa}_{j,\mathrm{s(ps)}}$ numerically
for $m=0$ and $m=0.1$
and find\footnote{We obtain 
$\sum_p|\tilde{\kappa}_{13,\mathrm{s}}(p)|^2/\tilde{\kappa}^2_{\rm sum,s}=1.000$
 and $\sum_p|\tilde{\kappa}_{24,\mathrm{ps}}(p)|^2/\tilde{\kappa}^2_{\rm sum,ps}=1.000$ with $\tilde{\kappa}^2_{\rm sum,s(ps)}=\sum_p\sum_{j=1,3,13,24,25,45}|\tilde{\kappa}_{j,\mathrm{s(ps)}}(p)|^2$, which shows that $\tilde{\kappa}_{13,\mathrm{s}}(p)$ and $\tilde{\kappa}_{24,\mathrm{ps}}(p)$ are dominant.}
\begin{align}
	\label{eq:ksstr}
	\tilde{S}^{(+)fg}_{\rm s}(p)
	\simeq
	\tilde{\kappa}_{13,\rm s}(p) \omega_{13}^{fg} \ ,  \quad \quad
	\tilde{S}^{(+)fg}_{\rm ps}(p)
	\simeq
	\tilde{\kappa}_{24,\rm ps}(p) \omega_{24}^{fg} \ .
\end{align}

Let us discuss the chiral transformation properties of
the Cooper pair condensate $\tilde{S}^{(+)fg}_{\rm s(ps)}(p)$. 
Here we focus on the $\mathrm{U}_{\rm c}(1)$ chiral symmetry of staggered fermions,
which is a remnant of the $\mathrm{SU}_{\rm L}(N_{\rm f})\times \mathrm{SU}_{\rm R}(N_{\rm f})$
chiral symmetry of the continuum theory defined by
the transformation
\begin{align}
	\label{eq:u1c_psi}
	\tilde{\psi}(p)\to e^{i\theta_{\rm c}\gamma_5 \otimes t_5}\tilde{\psi}(p) \ , 
	\quad \quad \theta_{\rm c} \in \mathbb{R} \ ,
\end{align}
where $\gamma_5=\gamma_1\gamma_2\gamma_3\gamma_4$ and $t_5$ act on the spinor
and flavor indices, respectively. It is straightforward to derive
the transformation
\begin{align}
	\tilde{S}^{(+)fg}_{\alpha\beta}(p)
	\to
	\tilde{S}^{(+)fg}_{\alpha\beta}(p)
	+
	i\theta_{\rm c}
	\left(
	\sum_{f',\alpha'}
	t_{5,ff'}\gamma_{5,\alpha\alpha'}
	\tilde{S}^{(+)f'g}_{\alpha'\beta}(p)
	+
	\sum_{g',\beta'}
	\tilde{S}^{(+)fg'}_{\alpha\beta'}(p)
	t_{5,gg'}\gamma_{5,\beta\beta'}
	\right)
\end{align}
under Eq.~\eqref{eq:u1c_psi} for infinitesimal $\theta_{\rm c}$.
By contracting the spinor indices as in Table~\ref{tab:lorentz} to extract
the scalar and pseudo-scalar components, one obtains the transformation
\begin{align}
	\tilde{S}_{\mathrm{s}}^{(+)fg}(p)
	\to&
	\tilde{S}_{\mathrm{s}}^{(+)fg}(p)
	+
	2i\theta_{\rm c} \tilde{\kappa}_{24,\rm ps}(p) \omega_{13}^{fg} \ ,
	\\
	\tilde{S}_{\mathrm{ps}}^{(+)fg}(p)
	\to&
	\tilde{S}_{\mathrm{ps}}^{(+)fg}(p)
	+
	2i\theta_{\rm c} \tilde{\kappa}_{13,\rm s}(p) \omega_{24}^{fg} \ ,
\end{align}
which amounts to
\begin{align}
	\begin{pmatrix}
		\tilde{\kappa}_{13,\rm s}(p)
	\\		
		\tilde{\kappa}_{24,\rm ps}(p)
	\end{pmatrix}
	\to
	e^{2i\theta_{\rm c} \sigma^{1}}
	\begin{pmatrix}
		\tilde{\kappa}_{13,\rm s}(p)
	\\		
		\tilde{\kappa}_{24,\rm ps}(p)
	\end{pmatrix}
\end{align}
for a finite $\theta_{\rm c}$.
Thus we find that the Copper pair condensate
breaks the $\mathrm{U}_{\rm c}(1)$ chiral symmetry spontaneously,
which is reflected in 
the double degeneracy for $m=0$ in Fig.~\ref{fig:beta_mqdep}.
A finite mass explicitly breaks the symmetry and
lifts the degeneracy of the scalar and pseudo-scalar condensates.

In the continuum limit, it is expected
that
the degeneracy of the largest eigenvalues in the massless case
enhances from 2 to 12 
due to the recovery of the
original
$\mathrm{SU}_{\rm L}(4)\times \mathrm{SU}_{\rm R}(4)$
chiral symmetry
since there are six ways to select two flavors for anti-commuting indices
from four flavors. In other words, all the eigenvalues corresponding to the twelve condensates $\tilde{\kappa}_{1,\mathrm{s(ps)}},\ldots,\tilde{\kappa}_{45,\mathrm{s(ps)}}$ in Eq.~\eqref{eq:Kdecomp} are expected to degenerate in the continuum limit,
which should be seen explicitly by using
larger lattices than the ones used in this work.

\subsection{Results for Wilson fermions \label{sec:res_wil}}

In this section we 
present our results for Wilson fermions,
which have the advantage of applicability
to any number of flavors
at the expense of the explicit chiral symmetry breaking.
The analysis based on the
gap equation is common to all $N_{\rm f} \geq 2$,
whereas the single flavor $N_{\rm f}=1$ case
has to be treated separately
since the absence of the flavor degrees of freedom
restricts the possible form of the condensate
due to the anti-commutating property of the fermion fields.
Here we first discuss the $N_{\rm f} \geq 2$ case comparing the results
with those for $N_{\rm f}=4$ staggered fermions,
and then discuss the $N_{\rm f}=1$ case.

\begin{figure}[t]
  \begin{center}
  	\includegraphics[width=\columnwidth]{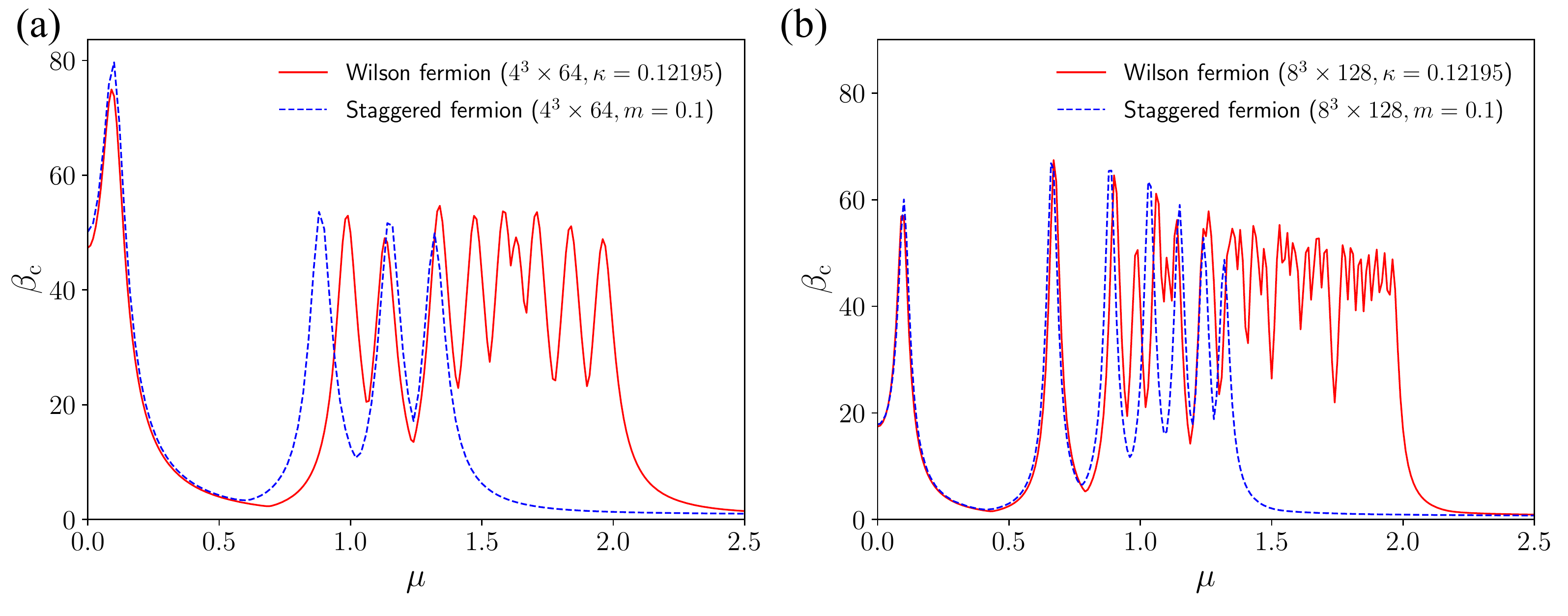}
        \caption{The critical coupling $\beta_c$ on (a) $4^3\times 64$ and (b) $8^3\times 128$ lattices
          is plotted against the quark chemical potential $\mu$
          for $N_{\rm f}\geq 2$
          Wilson (red solid lines) and staggered (blue dashed lines)
          fermions,
          respectively. 
          The hopping parameter for Wilson fermions is set to $\kappa=0.12195$,
          which corresponds to $m=0.1$ in the free theory.
          The mass of staggered fermions is set to $m=0.1$ for comparison.
        }
    \label{fig:wilson_staggered}
  \end{center}
\end{figure}
\begin{figure}[t]
  \begin{center}
    \includegraphics[width=0.75\columnwidth]{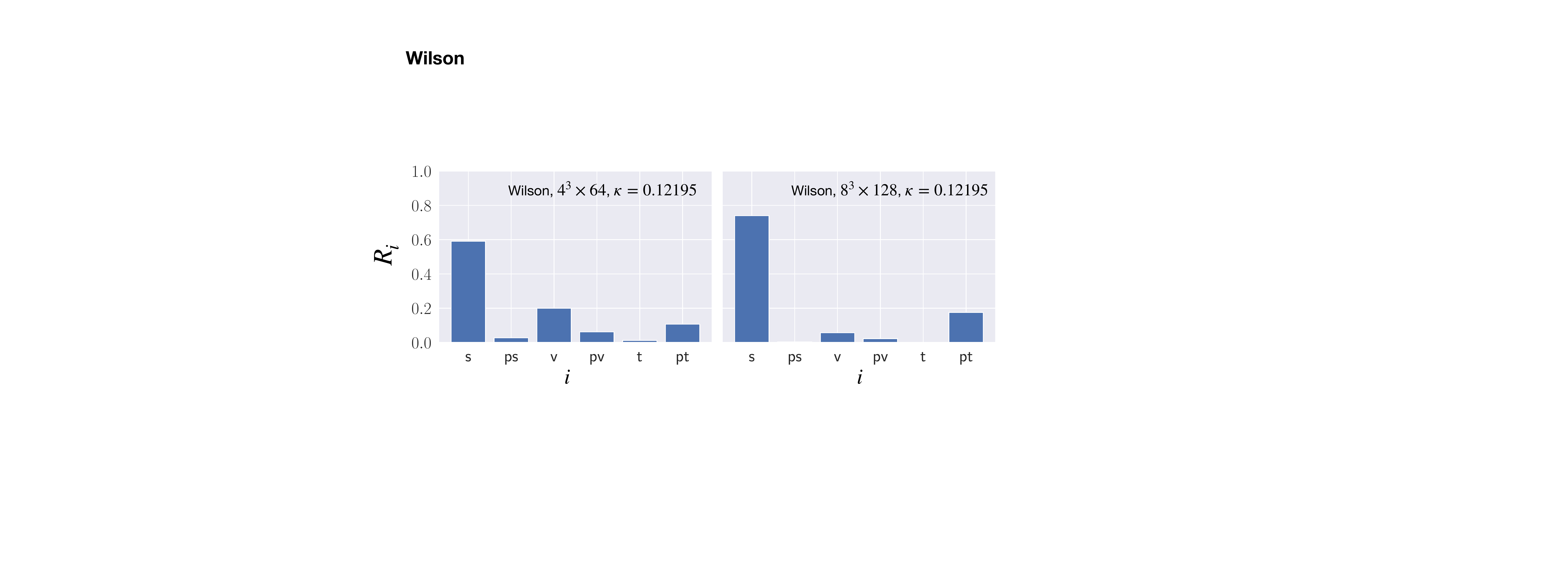}
        \caption{The components of the Cooper pair condensate
          $R_{i}$ ($i=\mathrm{s},\mathrm{ps}, \mathrm{v},\mathrm{pv}, \mathrm{t},\mathrm{pt}$)
          for $N_{\rm f}\geq 2$ Wilson fermions on $4^3\times 64$ and $8^3\times 128$ lattices
          with $\kappa=0.12195$ at $\mu = E(|{\bm p}| = 2 \pi / L_s)$.}
    \label{fig:strength_wilson}
  \end{center}
\end{figure}

Figure \ref{fig:wilson_staggered} shows the critical point $\beta_c$
as a function of the quark chemical potential $\mu$
for Wilson fermions together with the result for staggered fermions.
Note that the largest eigenvalue $\beta_c$ is non-degenerate.
Similarly to staggered fermions,
we observe 
a peak structure, where the peak positions correspond to the energy levels
$\mu=E({\bm p})$ with the dispersion relation
\begin{align}
	E({\bm p})=2\sinh^{-1}
	\sqrt{
	\frac{\sum_{i=1}^3\sin^2 p_i+\left(
	m+2\sum_{i=1}^3\sin^2\frac{p_i}{2}
	\right)^2
	}{4\left(1+m+2\sum_{i=1}^3\sin^2\frac{p_i}{2}
	\right)}}
	\ .
\end{align}
Here we define $m=1/(2\kappa)-4$ as
the quark mass in the free theory with the hopping parameter $\kappa$
and the Wilson parameter $r=1$.
The momentum ${\bm p}$ is chosen to be in 
the first Brillouin zone, which is given for Wilson fermions as
\begin{align}
	\label{eq:bz_wil}
	\lbrace
	{\bm p}|{\bm p}\in {\rm BZ_{\rm w}}
	\rbrace
	=&
	\left\lbrace
	\left.
	\left(
	\frac{2\pi n_1}{L_{s}},
	\frac{2\pi n_2}{L_{s}},
	\frac{2\pi n_3}{L_{s}}
	\right)
	\right|
	-\frac{L_{s}}{2}
	\leq n_{i}
	<\frac{L_{s}}{2},
	n_{i}\in \mathbb{Z}
	\right\rbrace.
\end{align}
The results for Wilson and staggered fermions agree in the small $\mu$ region,
which is understandable since
they have the same low-momentum properties
with the dispersion relation $E(\bm{p})\approx \sqrt{\bm{p}^2 +m^2}$.
Better agreement is observed for the larger lattice,
which shifts the peaks towards smaller $\mu$.
On the other hand, the results
exhibit
some discrepancies at large $\mu$.
Note, in particular, that the Wilson fermions have additional peaks there,
which is understood as a consequence
of the difference in the size of the first Brillouin zone
(See Eqs.~\eqref{eq:bz_staggered} and \eqref{eq:bz_wil}.).

In Fig.~\ref{fig:strength_wilson}
we show the components
of the Cooper pair condensate
by calculating $R_{i}$ ($i={\rm s},{\rm ps},{\rm v},{\rm pv},{\rm t},{\rm pt}$),
which are defined in the same manner as in the staggered fermion case
\eqref{eq:Rs}--\eqref{eq:Rt}.
We find that
the Cooper pair condensate
is of the scalar type, which agrees with the result for staggered
fermions with a finite mass.

As in the staggered fermion case, we calculate
\eqref{def-R-parity} and find
that  $R^{(+)}_{\rm s}\approx 0.9997$
for the scalar case with an $8^3 \times 128$ lattice,
which implies that the spatially symmetric component $\tilde{S}^{(+)fg}_{\rm s}(p)$
is dominant.
Thus we find that the dominant Cooper pair condensate is
anti-symmetric with respect to the flavor indices,
which implies two-flavor color superconductivity (2SC) for $N_{\rm f}=2$
and color-flavor locked color superconductivity (CFL) for $N_{\rm f}=3$.

\begin{figure}[!t]
  \begin{center}
  \begin{minipage}[b]{0.45\linewidth}
  	\centering
    \includegraphics[width=\columnwidth]{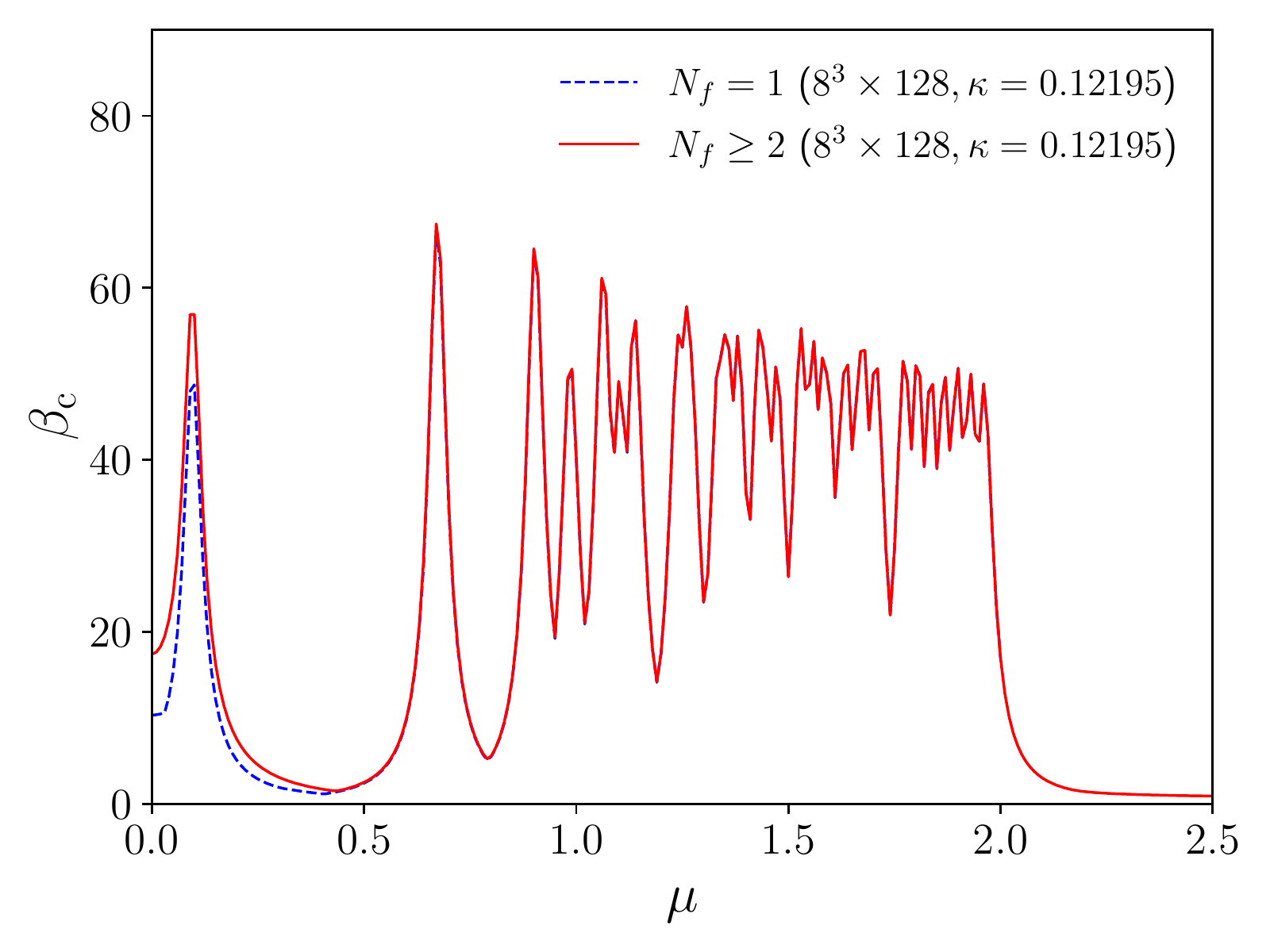}
  \end{minipage}
  \begin{minipage}[b]{0.45\linewidth}
  	\centering
    \includegraphics[width=\columnwidth]{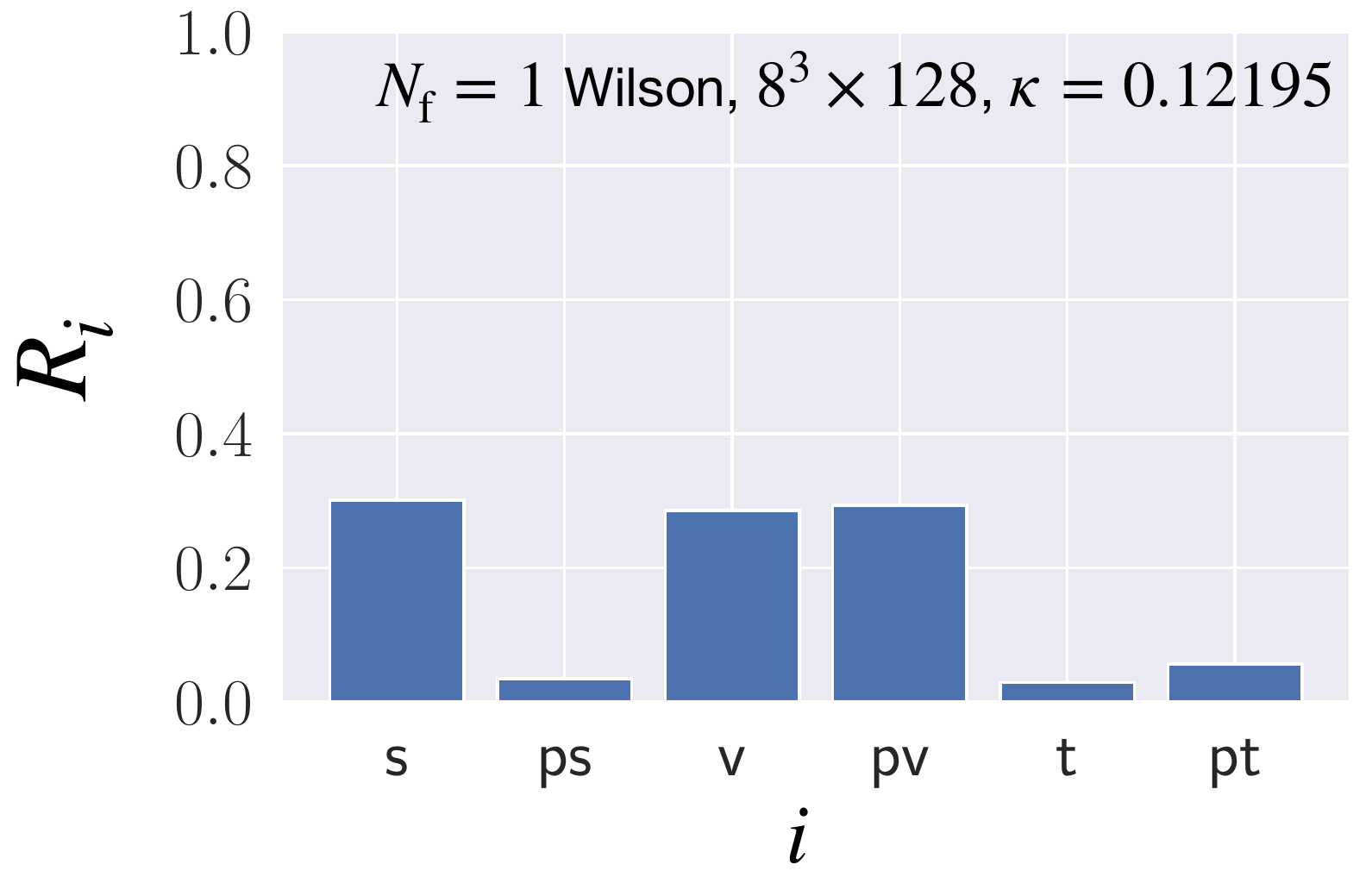}
  \end{minipage}
        \caption{(Left) The critical point $\beta_c$
          is plotted against the quark chemical potential
          for Wilson fermions
          on an $8^3\times 128$ lattice with $\kappa=0.12195$.
          The blue dashed line corresponds to the $N_{\rm f}=1$ case, whereas
          the red solid line corresponds to the $N_{\rm f}\geq 2$ case.
        (Right) The components of the Cooper pair condensate
          $R_{i}$ ($i=\mathrm{s},\mathrm{ps}, \mathrm{v},\mathrm{pv}, \mathrm{t},\mathrm{pt}$)
          for $N_{\rm f} = 1$ Wilson fermions on an $8^3\times 128$ lattice
          with $\kappa=0.12195$ and $\mu = E(|{\bm p}| = 2 \pi / L_s)$.
        }
    \label{fig:wilson_one_flavor}
  \end{center}
\end{figure}

Let us finally comment on the single-flavor case $N_{\rm f}=1$,
which is realized by restricting the eigenvector space so that the condensate
satisfies the anti-commuting property of the fermion fields
without the flavor indices as described in Appendices \ref{sec:m_form_wilson}
and \ref{sec:init}.
Figure~\ref{fig:wilson_one_flavor} shows a comparison
between $N_{\rm f} = 1$ and $N_{\rm f} \geq 2$.
From the left panel, we observe that the CSC region shrinks in the $N_{\rm f}=1$ case
due to the restriction of the eigenvector space.
From the right panel, we find that the scalar component is comparable
to the vector and pseudo-vector components unlike the $N_{\rm f}\geq 2$ case.
Note here that, in the case of $N_{\rm f}=1$,
the scalar condensate is spatially anti-symmetric suggesting the p-wave 
superconductivity, whereas the vector and pseudo-vector components are 
spatially symmetric suggesting the s-wave superconductivity. Which is 
realized in the continuum limit remains to be seen by calculations on a 
larger lattice. By the same token, the
difference of $\beta_c$ between the $N_{\rm f}=1$ and $N_{\rm f}\geq 2$ cases, which is visible 
only for $\mu \lesssim 0.4$ in the left panel, is expected to extend to the larger 
$\mu$ region in the continuum limit.

\section{Summary and discussions \label{sec:conc}}

In this paper we provided analytic predictions for the CSC on the lattice,
using the fact 
that the gap equation reduces to a linear equation
by focusing on the critical point.
In particular, we
determined the boundary of the normal and CSC phases
in the $\mu$-$\beta$ plane
for both staggered and Wilson fermions.
The phase boundary
shows characteristic peak structure as a function of the quark chemical potential,
which is
due to the discretization of the quark energy levels in finite systems.

Furthermore,
we investigated the form of the Cooper pair condensate
at the critical point.
In the case of staggered fermions,
we observed that the scalar and pseudo-scalar condensates are
favored in the massless limit of quarks
owing to the $\mathrm{U}_{\rm c}(1)$ chiral symmetry of staggered fermions,
which is a remnant of the $\mathrm{SU}_{\rm L}(4)\times \mathrm{SU}_{\rm R}(4)$ chiral symmetry
in the continuum.
The observed Cooper pair condensate
breaks the chiral symmetry spontaneously as well as
the $\mathrm{U}_{\rm B}(1)$ baryon-number symmetry.
When the quark mass is finite,
we found that
the degeneracy of the scalar and pseudo-scalar condensates is lifted
and that
the scalar condensate is favored.
From the momentum components of the condensate,
we confirmed that spatially isotropic s-wave superconductivity
is realized by the Cooper pairs composed of quarks near the Fermi surface.
As an extension of this work,
it would be interesting to include the chiral condensate,
which corresponds to the diagonal components of the Dyson equation
in the Nambu-Gor'kov formalism.

In the case of Wilson fermions, we find that 
the results for $N_{\rm f}\ge 2$ are essentially the same as staggered
fermions with a finite mass.
In particular, we find that the Cooper pair condensate is anti-symmetric
with respect to the flavor indices, which implies 2SC and CFL in
the case of $N_{\rm f}=2$ and $N_{\rm f}=3$, respectively.
The $N_{\rm f} = 1$ case was discussed separately and the results
turned out to be different from the $N_{\rm f}\ge 2$ case.

Our results obtained
for 
QCD in a small box
provides useful predictions 
for first-principle calculations based on methods to overcome the sign problem
such as the complex Langevin method.
Simulations in this direction are ongoing \cite{tsutsui_color_2022,namekawa_flavor_2022}.
Once our predictions are reproduced, the next step would be
to make the physical size of the box larger
either by decreasing $\beta$ or by using a larger lattice
in order to investigate the CSC in a fully non-perturbative regime.

\section*{Acknowledgements}
We thank Y.~Asano, E.~Itou, K.~Miura and A.~Ohnishi for valuable discussions.
T.~Y.~was supported by the RIKEN Special Postdoctoral Researchers Program. J
.~N.~was supported in part by JSPS KAKENHI Grant Numbers JP16H03988, JP22H01224.
Y.~N.~was supported by JSPS KAKENHI Grant Number JP21K03553. S.~T.~was supported by the RIKEN Special Postdoctoral Researchers Program.
Computations were carried out using computational resources of the Oakbridge-CX provided
by the Information Technology Center at the University of Tokyo through the HPCI System Research
project (Project ID:hp200079, hp210078, hp220094).
Computations were also carried out by using the computers in
the Yukawa Institute Computer Facility.

\appendix

\section{Explicit forms of $\mathcal{M}$ for staggered and
  Wilson fermions \label{sec:m_form}}

In this Appendix we give the explicit forms of $\mathcal{M}$ for
staggered and Wilson fermions.
We also make some remarks on the numerical calculation of $\mathcal{M}$.

\subsection{Staggered fermions \label{sec:m_form_staggered}}
Here we derive the form of $\mathcal{M}$ in the case of staggered fermions.
The action of staggered fermions with mass $m$ and the quark chemical potential $\mu$ in lattice units is given by
\begin{align}
	\label{eq:action}
	S & = 
	\frac{1}{2}\sum_{n,\dnu,a,a'}\eta_\dnu(n)
	\left\{
	\overline{\chi}^{a}(n)
	e^{\delta_{\dnu 4}
	\mu}
	U_{\dnu,aa'}(n)
	\chi^{a'}(n+\hat{\dnu})
	-
	\overline{\chi}^{a}(n-\hat{\dnu})
	e^{-\delta_{\dnu 4}\mu}
	U_{\dnu,aa'}^{\dagger}(n-\hat{\dnu})
	\chi^{a'}(n)
	\right\}
	\notag
	\\
	& \quad 
	+
	m\sum_{n,a}\overline{\chi}^{a}(n)\chi^{a}(n)
	+
	S_{\rm g} \ .
\end{align}
$S_{\rm g}$ is the action for gluons, $n$ is an integer vector
labeling the position on the hypercube, $a$ and $b$ are color indices,
$\hat{\dnu}$ is the unit vector in the $\dnu\ (=1,2,3,4)$ direction,
$\chi(n)$ and $\overline{\chi}(n)$ are the staggered fermion fields,
and $U_{\dnu}(n)=e^{ig\sum_{B}A_{\dnu}^{B}(n)T^{B}}$ is the link variable related with the gluon field $A_{\dnu}^{B}(n)$.
We have also introduced the usual
site-dependent sign factor $\eta_{\dnu}(n)=(-1)^{\sum_{\dnu'=1}^{\dnu-1}n_{\dnu'}}$.
We impose periodic boundary conditions
on $\chi(n)$ and $\overline{\chi}(n)$ in the spatial ($\dnu=1,2,3$)
directions and anti-periodic boundary conditions in the temporal ($\dnu=4$) direction.
The lattice extent in the $\dnu$ direction is denoted by $L_{\dnu}$.

First let us derive the Feynman rules for staggered fermions.
Following Ref.~\cite{rot12}, we redefine
the fermion field as
\begin{align}
	\label{eq:chi_rho}
	\chi_{\rho}^{a}(N)=\chi^{a}(2N+\rho) \ ,
\end{align}
where $\rho$ is a four-vector with $\rho_{\dnu}=0$ or $1$, while
$N$ is a new lattice coordinate labeling
the sites on a lattice with twice as large spacing as the original one.
This new field \eqref{eq:chi_rho}
is related to the four-flavor Dirac fermion field
through
\begin{align}
	\label{eq:dirac_staggered}
	\psi_{f\alpha}^{a}(N)
	\sim
	\sum_{\rho}
	(T_\rho)_{\alpha f}
	\chi_\rho^{a}(N) \ ,
\end{align}
where $\alpha$ and $f$ are the
spinor and flavor indices, respectively, and $T_\rho$ is defined by
\begin{align}
	T_{\rho}
	=
	\gamma_{1}^{\rho_1}
	\gamma_{2}^{\rho_2}
	\gamma_{3}^{\rho_3}
	\gamma_{4}^{\rho_4}
\end{align}
with $\gamma_{\mu}$ being the Euclidean Dirac gamma matrices.
In terms of
$\chi_\rho^{a}(N)$,
the free part of the action
is written as
\begin{align}
	\label{eq:action_free}
    S_{\rm f}^{\rm free}
    =&
    \sum_{N,N',\rho,\rho',a,a'}
    \overline{\chi}^{a}_\rho(N)
    D_{\rho\rho'}^{aa'}(N-N')
    \chi_{\rho'}^{a'}(N') \ ,
    \\
 	D_{\rho\rho'}^{aa'}(N)
	=&
	\delta_{a,a'}
    \sum_{\dnu}
    \frac{\eta_{\dnu}(\rho)}{2}
    \left\{
    e^{\delta_{\dnu 4}\mu}
    (
    \delta_{\rho+\hat{\dnu},\rho'}
    \delta_{N,0}
    +
    \delta_{\rho-\hat{\dnu},\rho'}
    \delta_{N+\hat{\dnu},0}
    )
    \right.
    \notag
    \\
	&-
    \left.
    e^{-\delta_{\dnu 4}\mu}
    (
    \delta_{\rho-\hat{\dnu},\rho'}
    \delta_{N,0}
    +
    \delta_{\rho+\hat{\dnu},\rho'}
    \delta_{N-\hat{\dnu},0}
    )
    \right\}
    +
    \tilde{m}_{\rm q}\delta_{a,a'}\delta_{\rho\rho'}\delta_{N,0} \ ,
\end{align}
where $\eta_\dnu(2N+\rho)=\eta_\dnu(\rho)$ has been used.
Let us switch to the momentum representation by using the Fourier transformation
\begin{align}
	\tilde{f}(p)
	=&
	\sum_{N}e^{-ip\cdot (2N)}f(N)
	\quad \quad 
	\text{for }f=\chi_{\rho}^a,\, \overline{\chi}_{\rho}^a  \ ,
\end{align}
and restrict the range of momentum $p$ to the first Brillouin zone
\begin{align}
	\label{eq:bz_f}
	\lbrace
	p|p\in {\rm BZ_{\rm s}}
	\rbrace
	=&
	\left\lbrace
	\left.
	\left(
	\frac{2\pi n_1}{L_1},
	\frac{2\pi n_2}{L_2},
	\frac{2\pi n_3}{L_3},
	\frac{(2n_4+1)\pi}{L_4}
	\right)
	\right|
	-\frac{L_{\dnu}}{4}
	\leq n_{\dnu}
	<\frac{L_{\dnu}}{4},
	n_{\dnu}\in \mathbb{Z}
	\right\rbrace \ ,
\end{align}
using the periodicity 
$\tilde{f}(p+\pi\hat{\dnu})=\tilde{f}(p)$.
Eq.~\eqref{eq:action_free} can then be written in the momentum representation as
\begin{align}
	\label{eq:action_free_momentum}
	S_{\rm f}^{\rm free}
    =&
    \sum_{p\in {\rm BZ}_{\rm s}}
    \sum_{\rho,\rho',a,a'}
    \tilde{\overline{\chi}}^{a}_\rho(p)
    \tilde{D}_{\rho\rho'}^{aa'}(p)
    \tilde{\chi}_{\rho'}^{a'}(-p) \ ,
    \\
	\label{eq:d}
    \tilde{D}_{\rho\rho'}^{aa'}(p)
    =&
    \delta_{aa'}
    \left(
    \sum_{\dnu}
    i\Gamma^\dnu_{\rho\rho'}(2p)
    \sin \overline{p}_{\dnu}
    +
    m\delta_{\rho\rho'}
    \right) \ ,
\end{align}
from which we obtain the fermion propagator
\begin{align}
	\tilde{D}_{\rho\rho'}^{-1,aa'}(p)
	=
	\delta_{aa'}
	\tilde{D}_{\rho\rho'}^{-1}(p)
	=
	\delta_{aa'}
	\frac{
    -\sum_{\dnu}
    i\Gamma^\dnu_{\rho\rho'}(2p)
    \sin \overline{p}_{\dnu}
    +
    m\delta_{\rho\rho'}
	}{
	\sum_\dnu \sin^2 \overline{p}_\dnu+m^2} \ .
\end{align}
Here we have introduced $\overline{p}_\dnu=p_\dnu -i \mu\delta_{\dnu 4}$ and
\begin{align}
	\label{eq:gamma}
	\Gamma^{\dnu}_{\rho\rho'}(2p)
	=e^{ip\cdot(\rho-\rho')} (
	\delta_{\rho+\hat{\dnu},\rho'}
	+\delta_{\rho-\hat{\dnu},\rho'})
	\eta_\dnu(\rho) \ ,
\end{align}
which satisfies the same algebra as the Dirac gamma matrices as
$\lbrace\Gamma^{\dnu}(2p),\Gamma^{\dnu'}(2p)\rbrace_{\rho\sigma}=2\delta_{\dnu\dnu'}\delta_{\rho\sigma}$.

By expanding $U_{\dnu}(n)=e^{ig\sum_{B}A_{\dnu}^{B}(n)T^{B}}$ with respect to $g$, we obtain
the interaction term
\begin{align}
	\label{eq:action_int}
    S_{\rm f}^{\rm int}
	=&
    ig\sum_{N,N',\rho,\rho',a,a',B}
    \overline{\chi}^{a}_\rho(N)
    \sum_{\dnu}
    \frac{\eta_{\dnu}(\rho)}{2}
    \notag
    \\
    &\times
    \left\{
    e^{\delta_{\dnu 4}\mu}
    A_{\dnu}^{B}(2N+\rho)T^{B}_{aa'}
    (
    \delta_{\rho+\hat{\dnu},\rho'}
    \delta_{N,N'}
    +
    \delta_{\rho-\hat{\dnu},\rho'}
    \delta_{N+\hat{\dnu},N'}
    )
    \right.
    \notag
    \\
	&-
    \left.
    e^{-\delta_{\dnu 4}\mu}
    A_{\dnu}^{B}(2N+\rho-\hat{\dnu})T^{B}_{aa'}
    (
    \delta_{\rho-\hat{\dnu},\rho'}
    \delta_{N,N'}
    +
    \delta_{\rho+\hat{\dnu},\rho'}
    \delta_{N-\hat{\dnu},N'}
    )
    \right\}
    \chi_{\rho'}^{a'}(N')
    +
    \mathcal{O}(g^2) \ ,
\end{align}
which can be rewritten
in the momentum space as
\begin{align}
    S_{\rm f}^{\rm int}
	=&
    \frac{1}{V'V}
    \sum_{k\in{\rm BZ}_{\rm g}}
    \sum_{p\in{\rm BZ}_{\rm s}}
    \sum_{\rho,\rho',a,a',B,\dnu}
    ig
	\Pi_{\rho\rho',aa'}^{B\dnu}(p,k)
    \tilde{\overline{\chi}}^a_{\rho}(-p)
    \tilde{A}_{\dnu}^{B}(-k)
    \tilde{\chi}^{a'}_{\rho'}(p+k) \ , 
\end{align}
using
the three-point vertex $\Pi_{\rho\rho',aa'}^{B\dnu}(p,k)$
given by
\begin{align}
	\label{eq:pi}
	\Pi_{\rho\rho',aa'}^{B\dnu}(p,k)
	=&
	e^{-ik\cdot \rho}
    \cos\left(
    \overline{p}_{\dnu}
    +
    \frac{{k}_{\dnu}}{2}
    \right)
    \Gamma^{\dnu}_{\rho\rho'}\left(2(p+k)\right)
	T^{B}_{aa'} \ .
\end{align}
Here we have introduced the momentum representation
for the gluon field \cite{rot12} as
\begin{align}
	A^B_{\dnu}(n)
	=
	\frac{1}{V}
	\sum_{k\in{\rm BZ}_{\rm g}}
	e^{ik\cdot n + ik_\dnu/2}\tilde{A}^B_{\dnu}(k) \ ,
\end{align}
where $V=L_1 L_2 L_3 L_4$ and
\begin{align}
	\lbrace
	k|k\in {\rm BZ}_{\rm g}
	\rbrace
	=
	\left\lbrace
        \left.
	\left(
	\frac{2\pi n_1}{L_1},
	\frac{2\pi n_2}{L_2},
	\frac{2\pi n_3}{L_3},
	\frac{2\pi n_4}{L_4}
	\right)
	\right|
	-\frac{L_{\dnu}}{2}
	\leq n_{\dnu}
	<\frac{L_{\dnu}}{2},
	n_{\dnu}\in \mathbb{Z}
	\right\rbrace.
\end{align}
Note that the range of momentum in the first Brillouin zone is different from
Eq.~\eqref{eq:bz_f} for the fermion field.
The gluon propagator is given by \cite{rot12}
\begin{align}
	G_{\dnu\dnu'}^{BB'}(k)
	=&
	\frac{\delta_{BB'}}{\tilde{k}^{2}}
	\left\{
	\delta_{\dnu\dnu'}
	-
	(1-\alpha_0)\frac{\tilde{k}_{\dnu}\tilde{k}_{\dnu'}}{\tilde{k}^{2}}
	\right\}
	=
	G(k)
	\delta_{BB'}
	\left\{
	\delta_{\dnu\dnu'}
	-
	(1-\alpha_0)\frac{\tilde{k}_{\dnu}\tilde{k}_{\dnu'}}{\tilde{k}^{2}}
	\right\} \ ,
\end{align}
where $\tilde{k}_\dnu=2\sin(k_\dnu/2)$ and $\alpha_0$ is the gauge parameter.
Since the results are independent of the
choice of $\alpha_0$ at the one-loop level,
we choose the Feynman gauge $\alpha_0=1$ for simplicity.

\begin{figure}[!t]
  \begin{center}
	\begin{align*}
	\parbox[c]{11em}{\includegraphics[width=11em]{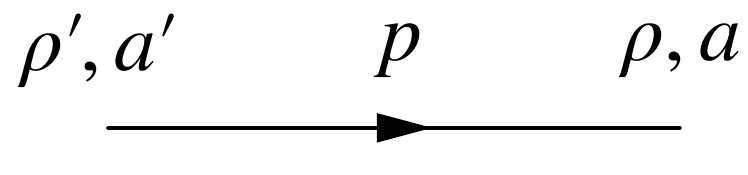}}
	&=
	\delta_{aa'}
	\tilde{D}_{\rho\rho'}^{-1}(p)
	=
	\delta_{aa'}
	\frac{
    -\sum_{\dnu}
    i\Gamma^\dnu_{\rho\rho'}(2p)
    \sin \overline{p}_{\dnu}
    +
    m\delta_{\rho\rho'}
	}{
	\sum_\dnu \sin^2 \overline{p}_\dnu+m^2}
	\\
	\parbox[c]{11em}{\includegraphics[width=11em]{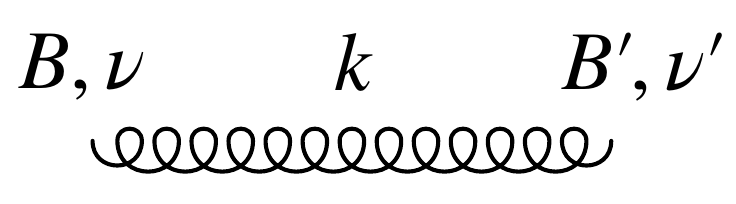}}
	&=
	\delta_{BB'}\delta_{\dnu\dnu'}
	G(k)
	=
	\delta_{BB'}\delta_{\dnu\dnu'}
	\frac{1}{4\sum_{\dnu''}\sin^2\left(\frac{k_{\dnu''}}{2}\right)}
	\\
	\parbox[c]{11em}{\includegraphics[width=11em]{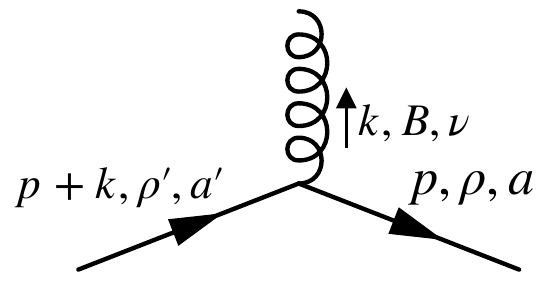}}
	&=
	igT_{aa'}^{B}\Pi_{\rho\rho'}^{\dnu}(p,k)
	=
	igT^{B}_{aa'}
	e^{-ik\cdot \rho}
    \cos\left(
    \overline{p}_{\dnu}
    +
    \frac{{k}_{\dnu}}{2}
    \right)
    \Gamma^{\dnu}_{\rho\rho'}\left(2(p+k)\right)
	\end{align*}
    \caption{Feynman rules for staggered fermions used in our calculation.}
    \label{fig:feynman_staggered}
  \end{center}
\end{figure}
\begin{figure}[!t]
  \begin{center}
  	\includegraphics[width=\columnwidth]{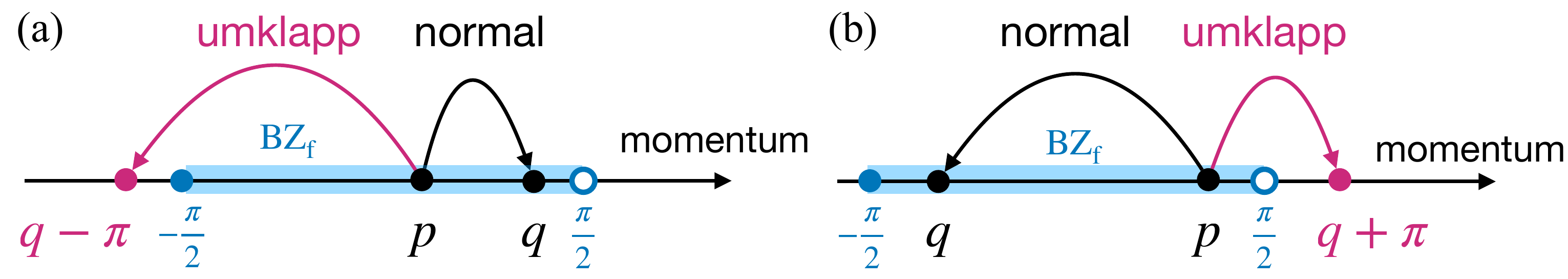}
        \caption{One-dimensional illustration of the two possible
          scatterings in the cases of (a) $p\leq q$ and (b) $q<p$
          for the fermion momenta $p$ and $q$.}
    \label{fig:umklapp}
  \end{center}
\end{figure}
Figure~\ref{fig:feynman_staggered} summarizes the Feynman rules
used in our calculation.
The other terms, such as the multi-gluon vertices,
contribute only to 
higher-order corrections and hence are ignored in our calculation.
Using these rules, we obtain $\mathcal{M}$ defined in Fig.~\ref{fig:sm_diag}(b)
as
\begin{align}
	\mathcal{M}_{(p\rho\rho')(q\sigma\sigma')}
	=&
	\frac{N_{\rm c}+1}{V}
	\sum_{\dnu,\gamma,\gamma'}
	\sum_{k \in {\rm BZ}_{\rm g}}
	\tilde{\delta}_{q-p-k}^{\rm F}
	G(k)
	\notag
	\\
	&\times
	\Pi_{\rho\gamma}^{\dnu}(p,k)
	\tilde{D}_{\gamma\sigma}^{-1}(p+k)
	\Pi_{\rho'\gamma'}^{\dnu}(-p,-k)
	\tilde{D}_{\gamma'\sigma'}^{-1}(-p-k) \ .
	\label{eq:Mstaggered}
\end{align}
Here we have introduced the periodic delta function
\begin{align}
	\tilde{\delta}_{k}^{\rm F}
	=
	\begin{cases}
		1 & (k_{1,\cdots,4}\in \pi\mathbb{Z})
		\\
		0 & (\text{otherwise}) \ ,
	\end{cases}
\end{align}
which
has the period $\pi$ in each direction inherited from
that
of $\mathrm{BZ}_{\mathrm{f}}$ in \eqref{eq:bz_f}.
It should be noted that the difference in the
size of the first Brillouin zone between fermions and gluons allows
two scattering processes, namely the normal and umklapp scatterings, in each
direction for any $p$ and $q$. Figure \ref{fig:umklapp} is a one-dimensional
illustration of these processes. Due to the range of the gluon
momentum $-\pi\leq k< \pi$, not only the normal process but also the
umklapp process, where $p+k$ exceeds the domain of the first Brillouin zone,
always occurs for any fermion momenta. In four dimensions, the existence of both scatterings in each direction is represented by the
solution of $\tilde{\delta}_{q-p-k}^{\rm F}=1$ in $-\pi\leq k_\dnu< \pi$
given by
\begin{align}
	\label{eq:ksol}
	k_\dnu
	=&
	q_\dnu-p_\dnu - \pi\rho''_{\dnu} \, {\rm sgn}_{+}(q_\dnu-p_\dnu)
	\quad (\rho''_{\dnu}=0,1)
\end{align}
with
\begin{align}
	{\rm sgn}_{+}(x)
	=
	\begin{cases}
		1 & \mbox{~for~}x\geq 0
		\\
		-1 &  \mbox{~for~} x< 0 \ .
	\end{cases}
\end{align}
After performing the sum over $k$ in Eq.~\eqref{eq:Mstaggered}, we obtain
\begin{align}
	\label{eq:M_BZreduced}
	\mathcal{M}_{(p\rho\rho')(q\sigma\sigma')}
	=&
	\frac{N_{\rm c}+1}{V}
	\sum_{\dnu,\gamma,\gamma'}
	\sum_{\rho''}
	G\left(k(q,p,\rho'')\right)
	\Pi_{\rho\gamma}^{\dnu}\left(p,k(q,p,\rho'')\right)
	\tilde{D}_{\gamma\sigma}^{-1}(p+k(q,p,\rho''))
	\notag
	\\
	&\times
	\Pi_{\rho'\gamma'}^{\dnu}\left(-p,-k(q,p,\rho'')\right)
	\tilde{D}_{\gamma'\sigma'}^{-1}(-p-k(q,p,\rho'')) \ ,
\end{align}
where $k(q,p,\rho'')$ is defined by Eq.~\eqref{eq:ksol}.
In the case of Wilson fermions discussed in the following subsection,
only one of the normal and umklapp scatterings occurs for a given momentum
transfer because there is no difference in the first Brillouin zone between
fermions and gluons.

\subsection{Wilson fermions
  \label{sec:m_form_wilson}
}

Next let us
derive the form of $\mathcal{M}$ in the case of Wilson fermions,
where the fermion fields on the lattice site $N$ are
denoted by $\psi^{a}_{f\alpha}(N)$ and $\overline{\psi}^{a}_{f\alpha}(N)$
with $a$, $f$ and $\alpha$ being the color,
flavor and spinor indices, respectively. The action is given by
\begin{align}
	\label{eq:action_wilson}
    S
    =&
	\frac{1}{2}
    \sum_{N,a,a',\alpha,\alpha',f,\dnu}
	\left\{
    \overline{\psi}^{a}_{f\alpha}(N)
	(\gamma_{\dnu})_{\alpha\alpha'}
	e^{\mu\delta_{\dnu 4}}
	U_{\dnu,aa'}(N)
    \psi^{a'}_{f\alpha'}(N+\hat{\dnu})
	\right.
	\notag
	\\
	&\left.
	\qquad\qquad
	\qquad\,
	-
    \overline{\psi}^{a}_{f\alpha}(N+\hat{\dnu})
	(\gamma_{\dnu})_{\alpha\alpha'}
	e^{-\mu\delta_{\dnu 4}}
	U_{\dnu,aa'}^\dagger(N)
    \psi^{a'}_{f\alpha'}(N)
	\right\}
	+
    m
    \sum_{N,a,\alpha,f}\overline{\psi}^{a}_{f\alpha}(N)\psi_{f\alpha}^{a}(N)
    \notag
    \\
    &
	-\frac{r}{2}
    \sum_{N,a,a',\alpha,f,\dnu}
	\left\{
    \overline{\psi}^{a}_{f\alpha}(N)
	e^{\mu\delta_{\dnu 4}}
	U_{\dnu,aa'}(N)
    \psi^{a'}_{f\alpha}(N+\hat{\dnu})
	\right.
	\notag
	\\
	&\left.
	\qquad\qquad
	\qquad\,
	-
	2
    \overline{\psi}^{a}_{f\alpha}(N)
    \delta_{a,a'}
	\psi^{a'}_{f\alpha}(N)
	+
    \overline{\psi}^{a}_{f\alpha}(N+\hat{\dnu})
	e^{-\mu\delta_{\dnu 4}}
	U_{\dnu,aa'}^\dagger(N)
    \psi^{a'}_{f\alpha}(N)
	\right\}
    +
    S_{\rm g} \ ,
\end{align}
where $m$ and $\mu$ are the mass and the quark chemical potential in lattice units,
respectively, and $r$ is the Wilson parameter, which defines the hopping parameter as $\kappa=1/(2m+8r)$.

\begin{figure}[t]
  \begin{center}
	\begin{align*}
	\parbox[c]{11em}{\includegraphics[width=11em]{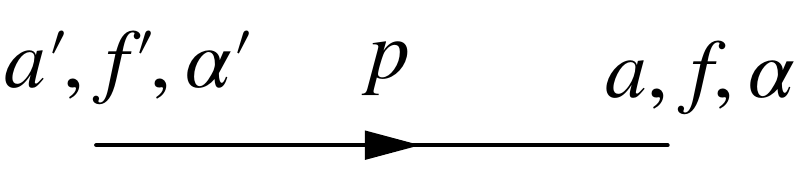}}
	&=
	\delta_{aa'}
	\delta_{ff'}
	\tilde{D}_{\alpha \alpha'}^{-1}(p)
	=
	\delta_{aa'}
	\delta_{ff'}
	\frac{
    -\sum_{\dnu}
    i\gamma^\dnu_{\alpha\alpha'}
    \sin \overline{p}_{\dnu}
    +
    M(\overline{p})\delta_{\alpha\alpha'}
	}{
	\sum_\dnu \sin^2 \overline{p}_\dnu+M(\overline{p})^2}
	\\
	\parbox[c]{11em}{\includegraphics[width=11em]{fig_gluon}}
	&=
	\delta_{BB'}\delta_{\dnu\dnu'}
	G(k)
	=
	\delta_{BB'}\delta_{\nu\nu'}
	\frac{1}{4\sum_{\dnu''}\sin^2\left(\frac{k_{\dnu''}}{2}\right)}
	\\
	\parbox[c]{11em}{\includegraphics[width=11em]{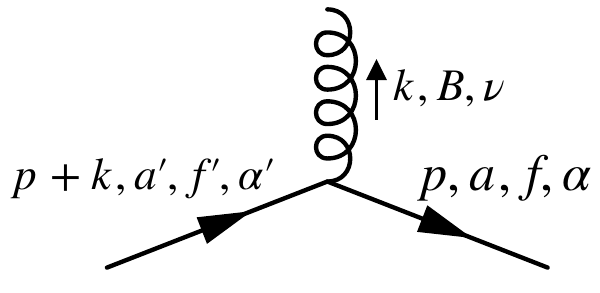}}
	&=
	-ig\delta_{ff'}
	T^{B}_{aa'}
	\Pi^{\dnu}_{\alpha\alpha'}(q,q')
	\\
	&=
	-ig\delta_{ff'}
	T^{B}_{aa'}
	\left(	
	\gamma_{\dnu,\alpha\alpha'}
	\cos \frac{\overline{q}_{\dnu}^{'}-\overline{q}_{\dnu}^{*}}{2}
	-
	\delta_{\alpha\alpha'}
	ir
	\sin \frac{\overline{q}_{\dnu}^{'}-\overline{q}_{\dnu}^{*}}{2}
	\right)
	\end{align*}
        \caption{Feynman rules for Wilson fermions used in our calculation.
          We have introduced $\overline{q}_\nu=q_\nu-i\mu\delta_{\nu 4}$
          and $M(\overline{p})=m+2r\sum_\nu\sin^2(\overline{p}_\nu/2)$.}
    \label{fig:feynman_wilson}
  \end{center}
\end{figure}
Similarly to staggered fermions, the Feynman rules are derived from
the action \cite{rot12}. In Fig.~\ref{fig:feynman_wilson} we summarize the Feynman rules in the momentum space used in our calculation. According to these
rules, $\mathcal{M}$ is evaluated as
\begin{align}
	\label{eq:m_wilson}
	\mathcal{M}_{(pf\alpha f'\alpha'),(q h\beta h'\beta')}
	=&
	\delta_{fh}\delta_{f'h'}
	\frac{N_{\rm c}+1}{V}
	G(q-p)
	\sum_{\gamma,\gamma',\mu}
	\Pi_{\alpha'\gamma'}^{\mu}(-p,q)
	\tilde{D}^{-1}_{\gamma'\beta'}(q)
	\Pi_{\alpha\gamma}^{\mu}(p,-q)
	\tilde{D}^{-1}_{\gamma\beta}(-q)
\end{align}
with $V=L_1 L_2 L_3 L_4$. As for the indices of $\mathcal{M}$, let us recall
that the indices $\rho$, $\rho'$, $\sigma$ and $\sigma'$ used in
Sec.~\ref{sec:form}
represent the flavor and spinor indices collectively.
The first Brillouin zone for Wilson fermions
is given by the momentum region
\begin{align}
	\lbrace
	p|p\in {\rm BZ_{\rm w}}
	\rbrace
	=
	\left\lbrace
        \left.
	\left(
	\frac{2\pi n_1}{L_1},
	\frac{2\pi n_2}{L_2},
	\frac{2\pi n_3}{L_3},
	\frac{2\pi n_4}{L_4}
	\right)
	\right|
	-\frac{L_{\dnu}}{2}
	\leq n_{\dnu}
	<\frac{L_{\dnu}}{2},
	n_{\dnu}\in \mathbb{Z}
	\right\rbrace \ .
\end{align}

Since $\mathcal{M}$ trivially acts on the flavor indices,
i.e., $\mathcal{M}_{(pf\alpha f'\alpha'),(q h\beta h'\beta')}\propto \delta_{fh}\delta_{f'h'}$,
the largest eigenvalue is independent
of $N_{\rm f}$
as far as the number of flavors is $N_{\rm f}\ge 2$.
For the single flavor case $N_{\rm f}=1$, however,
the largest eigenvalue can be different
because of the restriction on the eigenvectors
$\tilde{\Sigma}^{(-)}_{12(pf\alpha f'\alpha')}$
due to the anti-commuting
property of the fermion fields (See Appendix \ref{sec:init}.).
Namely, since
$\tilde{\Sigma}^{(-)}_{12(pf\alpha f'\alpha')}=\tilde{\Sigma}^{(-)}_{12(p\alpha\alpha')}$,
the eigenvector must satisfy
$\tilde{\Sigma}^{(-)}_{12(p\alpha\alpha')}=\tilde{\Sigma}^{(-)}_{12(-p\alpha'\alpha)}$
because of Eq.~\eqref{eq:sigma_sym}.
This is in contrast to the multi-flavor case, where more general
types of condensate are allowed since $\tilde{\Sigma}^{(-)}_{12(pf\alpha f'\alpha')}$
can be anti-symmetric with respect to the flavor indices $f$ and $f'$.

\subsection{Some remarks on the calculation of $\mathcal{M}$}

Here we make some remarks on
the numerical calculation of $\mathcal{M}$ that applies to both staggered
and Wilson fermions.
First we point out that
the elements corresponding to zero momentum transfer diverge
due to $G(0)$, which is the contribution from the gluon zero modes.
This is an artifact of perturbation theory on a finite lattice,
which disappears in the large volume limit.
Note also that the divergence does not appear in
non-perturbative treatments (See Ref.~\cite{Asano:2023nef}, for instance.).
In this work, we simply omit the contribution from $G(0)$ by hand.

The next point concerns the memory consumption by $\mathcal{M}$, which is
a huge and dense matrix. For instance, for staggered fermions, the number of
elements is $256V^2$ for the lattice volume $V$, which amounts
to $18{\rm TB}$ for $V=8^3\times 128$
with double precision complex numbers. Therefore, keeping all the elements in
the memory is not practical.
Instead, we decompose $\mathcal{M}$ into
parts whose memory consumption is at most $O(V)$, such as $G$ and
$\tilde{D}^{-1}$ in Eq.~\eqref{eq:M_BZreduced}, and calculate the elements of
$\mathcal{M}$ from these parts every time they are needed.

\section{Details of the power iteration method
  \label{sec:power-iter}
}

\subsection{Initial condition for the power iteration\label{sec:init}}
In the power iteration method,
one extracts the largest eigenvalue and the corresponding eigenvector
by multiplying $\mathcal{M}$ many times to a randomly selected initial vector.
However, in order to obtain eigenvectors with appropriate symmetry properties,
one needs to impose some condition
on the initial vector.

Note that the anomalous propagator for fermions 
$\tilde{S}_{12,\rho\rho'}^{(aa')}(p)=\langle \tilde{\psi}^{a}_{\rho}(p)\tilde{\psi}^{a'}_{\rho'}(-p) \rangle$
satisfies the
relation
\begin{align}
	\tilde{S}_{12,\rho\rho'}^{aa'}(p)
	=
	-\tilde{S}_{12,\rho'\rho}^{a'a}(-p) \ . 
\end{align}
Similarly, the anomalous self-energy satisfies
\begin{align}
	\label{eq:anti_com}
	\tilde{\Sigma}_{12,\rho\rho'}^{aa'}(p)
	=
	-\tilde{\Sigma}_{12,\rho'\rho}^{a'a}(-p) \ ,
\end{align}
as one can see from Eq.~\eqref{eq:dyson}. Therefore, we have
\begin{align}
	\label{eq:sigma_sym}
	\tilde{\Sigma}_{12(p\rho\rho')}^{(-)}
	=
	\tilde{\Sigma}_{12(-p\rho'\rho)}^{(-)}
\end{align}
for $\tilde{\Sigma}_{12(p\rho\rho')}^{(-)}$ in Eq.~\eqref{eq:gap_linear}. 

Let us decompose the vector space $\mathcal{V}$ on which $\mathcal{M}$ acts
as
$\mathcal{V}=\mathcal{V}_{\rm F}+\mathcal{V}_{\rm B}$,
where $\mathcal{V}_{\rm F(B)}$ is the vector space whose elements satisfy
\begin{alignat}{3}
	\label{eq:vf_def}
	v_{(q\rho\rho')}=&v_{(-q\rho'\rho)}
	\quad
	&\text{for }v\in \mathcal{V}_{\rm F} \ ,
	\\
	v_{(q\rho\rho')}=&-v_{(-q\rho'\rho)}
	\quad
	&\text{for }v\in \mathcal{V}_{\rm B} \ .
\end{alignat}
By using the relation 
\begin{align}
	\mathcal{M}_{(p\rho\rho')(q\sigma\sigma')}
	=
	\mathcal{M}_{(-p\rho'\rho)(-q\sigma'\sigma)} \ ,
\end{align}
which is obtained from the representation in Fig.~\ref{fig:sm_diag}(b),
one can show that
\begin{alignat}{3}
	\label{eq:mv_f}
	\mathcal{M}v \in \mathcal{V}_{\rm F}
	\quad
	&\text{if }v\in \mathcal{V}_{\rm F} \ ,
	\\
	\label{eq:mv_b}
	\mathcal{M}v \in \mathcal{V}_{\rm B}
	\quad
	&\text{if }v\in \mathcal{V}_{\rm B} \ ,
\end{alignat}
which implies that $\mathcal{V}_{\rm F}$ and $\mathcal{V}_{\rm B}$ are not mixed by multiplying $\mathcal{M}$.
Therefore, the initial vector must satisfy Eq.~\eqref{eq:vf_def} in order to obtain the eigenvector satisfying Eq.~\eqref{eq:sigma_sym}.

The eigenvalue equation for Wilson fermions reduces to
\begin{align}
	\label{eq:m_wilson_reduced}
	\sum_{q\beta\beta'}
	\mathcal{M}_{(p\alpha\alpha')(q\beta\beta')}'
	\tilde{\Sigma}^{(-)}_{12(qf\beta f'\beta')}
	=
	\beta \, 
	\tilde{\Sigma}^{(-)}_{12(pf\alpha f'\alpha')}
        \ ,
\end{align}
where $\mathcal{M}'$ is related to $\mathcal{M}$ given by
Eq.~\eqref{eq:m_wilson} as
\begin{align*}
	\mathcal{M}_{(pf\alpha f'\alpha'),(q h\beta h'\beta')}
	=
	\delta_{fh}\delta_{f'h'}	\mathcal{M}_{(p\alpha \alpha'),(q \beta \beta')}'
        \ .
\end{align*}
As we mentioned in Appendix \ref{sec:m_form_wilson},
the allowed forms of the eigenvector $\tilde{\Sigma}^{(-)}_{12(pf\alpha f'\alpha')}$
are different between the single- and multi-flavor cases for Wilson fermions.
Let us decompose
$\tilde{\Sigma}^{(-)}_{12(pf\alpha f'\alpha')}$
into the symmetric and antisymmetric
parts with respect to $f$ and $f'$. Abbreviating the flavor indices, we denote
the former and latter parts by
$\tilde{\Sigma}^{(-+)}_{12(p\alpha \alpha')}$ and
$\tilde{\Sigma}^{(--)}_{12(p\alpha \alpha')}$, respectively.
In the single-flavor
case, $\tilde{\Sigma}^{(--)}_{12(q\beta \beta')}$ does not exist, which forces us
to
choose the initial vector to satisfy the condition
\begin{align}
	\tilde{\Sigma}^{(-+)}_{12(q\beta \beta')}
	&=
	\tilde{\Sigma}^{(-+)}_{12(-q\beta' \beta)}
\end{align}
due to Eq.~\eqref{eq:sigma_sym}.

\subsection{Extension
    to the second and the third largest eigenvalues \label{sec:method_eigens}}
In this section we extend the power iteration method to
the calculation of
the second and the third largest
eigenvalues as well as the corresponding eigenvectors.
Let us assume that the largest eigenvalue $\lambda_1$
and the corresponding eigenvector $v_1$
of $\mathcal{M}$ defined in Fig.~\ref{fig:sm_diag}(b)
are obtained.
If one defines $\tilde{v}$ by projecting out
the $v_1$ component from $v$ and applies
the power iteration to $\tilde{v}$, the second largest eigenvalue
$\lambda_2$ and the corresponding eigenvector $v_2$ are obtained.
By repeating this procedure, one obtains the third largest one as well.

If $\mathcal{M}$ were Hermitian,
the projection can be made
by using the orthogonality of the eigenvectors.
In fact, $\mathcal{M}$ is not Hermitian
but pseudo-Hermitian \cite{dir42,pauli_diracs_1943,lee_negative_1969}
\begin{align}
	\label{eq:lml}
	\mathcal{M}_{(p\rho\rho')(q\sigma\sigma')}^{\dagger}
	=
	\left[
	\eta\mathcal{M}\eta^{-1}
	\right]_{(p\rho\rho')(q\sigma\sigma')}
\end{align}
with a Hermitian matrix $\eta$
for both staggered and Wilson fermions.

In order to make the projection in this case,
we need to consider the relationship
among the eigenvectors under the condition \eqref{eq:lml}. Let $\lambda_n$
and $v_n$ be an eigenvalue and the corresponding eigenvector, respectively,
with the ordering
$|\lambda_1|\geq |\lambda_2|\geq \cdots$. The eigenvalue equation reads
\begin{align}
	\label{eq:Mvn}
	\mathcal{M}v_n=&\lambda_n v_n \ .
\end{align}
By using its Hermitian conjugate and Eq.~\eqref{eq:lml}, we obtain
\begin{align}
	v_n^\dagger \eta\mathcal{M}=\lambda_n^{*} v_n^{\dagger}\eta \ .
\end{align}
Acting this on $v_m$ and using Eq.~\eqref{eq:Mvn}, we have
\begin{align}
	(\lambda_m-\lambda_n^{*})v_n^\dagger \eta v_m=0 \ .
\end{align}
From this, we find that
the eigenvalue $\lambda_n$ is real 
if $v_n^\dagger \eta v_n\neq 0$
and that the eigenvectors satisfy
$v_n^\dagger \eta v_m=0$ if $\lambda_m\neq \lambda_n^{*}$,
which can be regarded as a generalization of the properties
in the Hermitian case.

Suppose we have already obtained $v_{1},v_{2},\ldots,v_{m-1}$ for an integer
$m\geq 1$, which satisfy $v_n^\dagger \eta v_n\neq 0$ and
$v_{n}^\dagger \eta v_{n'}=0$ for all $n,n'<m$.
Then we can get rid of these components as
\begin{align}
	\label{eq:vmap}
	\tilde{v}
	=
	v
	-
	\sum_{n < m}
           a_n
	v_n \ ,
\end{align}
where the coefficient $a_n$ for $n< m$ is given by
\begin{align}
	a_n
	=
	\frac{v^{\dagger}_n \eta v}{v^{\dagger}_n \eta v_n} \ .
\end{align}
As mentioned above, we can extract $\lambda_m$ and $v_m$ by applying the
power iteration to $\tilde{v}$. As far as $v_n^\dagger \eta v_n\neq 0$ is
satisfied for the obtained eigenvectors, we can repeat the procedure to
extract other eigenvalues and eigenvectors. In the process of our calculations,
we checked $v_n^\dagger \eta v_n\neq 0$ for the obtained eigenvectors.

\providecommand{\href}[2]{#2}\begingroup\raggedright\endgroup

\end{document}